\documentclass[journal]{IEEEtran}
\newcommand{\CLASSINPUTtoptextmargin}{2.4cm}
\usepackage{graphicx}
\usepackage{subfigure}
\usepackage{float}
\usepackage{amsmath}
\usepackage{amsfonts,amssymb}
\usepackage{dsfont}
\usepackage{algpseudocode}
\usepackage{makecell}
\usepackage{cite}
\usepackage{multirow}
\usepackage{multicol}
\usepackage{booktabs}
\usepackage[linesnumbered, lined, ruled, noend]{algorithm2e}
\usepackage{soul}
\usepackage{tabularx}
\usepackage[table,xcdraw]{xcolor}
\usepackage{rotating}
\usepackage{afterpage}
\usepackage{pdflscape}
\usepackage{url}
\usepackage[english]{babel}
\usepackage[autostyle]{csquotes}
\usepackage{bbm}
\usepackage[flushleft]{threeparttable}
\usepackage{xcolor}

\newcommand{\mycomment}[1]{}
\newcommand*{\myprime}{^{\mkern 1.2mu \prime}}

\newcolumntype{C}[1]{>{\arraybackslash}p{#1}}

\newcommand{\myHighlight}[1]{{#1}}
\newcommand{\secH}[1]{{#1}}
\definecolor{myhl}{RGB}{255,242,11}

\SetCommentSty{mycommfont}

\SetKwInput{KwInput}{Input}
\SetKwInput{KwOutput}{Output}
\def\BibTeX{{\rm B\kern-.05em{\sc i\kern-.025em b}\kern-.08em
    T\kern-.1667em\lower.7ex\hbox{E}\kern-.125emX}}

\DeclareFontFamily{U}{mathx}{}
\DeclareFontShape{U}{mathx}{m}{n}{<-> mathx10}{}
\DeclareSymbolFont{mathx}{U}{mathx}{m}{n}
\DeclareMathAccent{\widehat}{0}{mathx}{"70}
\DeclareMathAccent{\widecheck}{0}{mathx}{"71}

\DeclareMathAlphabet{\mathpzc}{OT1}{pzc}{m}{it}

\begin{document}

\title{Energy Efficient Orchestration in Multiple-Access Vehicular Aerial-Terrestrial 6G Networks}

\author{
    \IEEEauthorblockN{
        Mohammad Farhoudi\textsuperscript{1}, Hamidreza Mazandarani\textsuperscript{2}, Masoud Shokrnezhad\textsuperscript{3}, Tarik Taleb\textsuperscript{2}, and Ignacio Lacalle\textsuperscript{4} \\
    }
    \IEEEauthorblockA{
        \textsuperscript{1} \textit{Oulu University, Finland};
        mohammad.farhoudi@oulu.fi \\
        \textsuperscript{2} \textit{Ruhr University Bochum (RUB), Germany}; \{hamidreza.mazandarani, tarik.taleb\}@rub.de \\ 
        \textsuperscript{3} \textit{ICTFICIAL Oy, Espoo, Finland}; masoud.shokrnezhad@ictficial.com \\
        \textsuperscript{4} \textit{Universitat Politècnica de València, Spain}; iglaub@upv.es
        \vspace{-10pt}
    }
}

\maketitle

\IEEEpubid{%
\parbox{\textwidth}{\centering\footnotesize
The manuscript is accepted for publication in IEEE Transactions on Vehicular Technology. Copyright\copyright~2026 IEEE. Personal use of this material is permitted. However, permission to use this material for any other purposes must be obtained from the IEEE by sending a request to pubs-permissions@ieee.org.
}%
}
\IEEEpubidadjcol
\IEEEpubidadjcol

\begin{abstract}
The proliferation of users, devices, and novel vehicular applications--propelled by advancements in autonomous systems and connected technologies--is precipitating an unprecedented surge in novel services. These emerging services require substantial bandwidth allocation, adherence to stringent Quality of Service (QoS) parameters, and energy-efficient implementations, particularly within highly dynamic vehicular environments. The complexity of these requirements necessitates a fundamental paradigm shift in service orchestration methodologies to facilitate seamless and robust service delivery. This paper addresses this challenge by presenting a novel framework for service orchestration in Unmanned Aerial Vehicles (UAV)-assisted 6G aerial-terrestrial networks. The proposed framework synergistically integrates UAV trajectory planning, Multiple-Access Control (MAC), and service placement to facilitate energy-efficient service coverage while maintaining ultra-low latency communication for vehicular user service requests. We first present a non-linear programming model that formulates the optimization problem. Next, to address the problem, we employ a Hierarchical Deep Reinforcement Learning (HDRL) algorithm that dynamically predicts service requests, user mobility, and channel conditions, addressing the challenges of interference, resource scarcity, and mobility in heterogeneous networks. Simulation results demonstrate that the proposed framework outperforms state-of-the-art solutions in request acceptance, energy efficiency, and latency minimization, showcasing its potential to support the high demands of next-generation vehicular networks.
\end{abstract}


\section{Introduction}
The rapid proliferation of connected autonomous vehicles and associated devices, such as on-board units and short-range communication transceivers, is driving unprecedented growth in both technological sophistication and deployment volume. 
Technological advancements have enabled vehicular User Equipment (UEs) to integrate with diverse vehicular services, including object detection, traffic analysis, and high-precision cartographic updates \cite{chen2020decentralized}, playing a pivotal role in augmenting vehicular functionality and enhancing it in various aspects like road safety protocols \cite{DqnResourceAlloc}.
However, the exponential growth in deployment density has precipitated unprecedented traffic demands \cite{6GOrchestration}, a phenomenon amplified by the cumulative effect of each additional vehicle transmitting increasingly data-intensive sensor readings, high-definition video streams, and telemetry information simultaneously \cite{PathMng}. This multifaceted data proliferation renders the provision of continuous service access for UEs a critical technical challenge \cite{V2XChallenge}.

Vehicular services are architected through the integration of multiple fundamental functions, each executing a discrete task. For instance, real-time traffic monitoring represents a paradigmatic composed service, synthesizing distinct functional components including vehicle velocity monitoring, safety-critical message dissemination, and traffic density quantification. These constituent functions operate in parallel and synergistically to deliver comprehensive service functionality. Such composition necessitates adherence to stringent Quality of Service (QoS) parameters, with ultra-low End-to-End (E2E) latency emerging as the predominant requirement for real-time communications \cite{wright_autonomous_cars, federateMultiAgent}. These exacting performance criteria present significant implementation challenges, as contemporary service orchestration approaches demonstrate inadequate capability to consistently maintain uninterrupted connectivity while satisfying the requisite E2E latency thresholds \cite{Fair5GV2x}.

The evolution of networks has enabled advanced solutions tailored to emerging vehicular services, among which the vehicular edge-cloud continuum has emerged as a promising paradigm for real-time vehicular services. By leveraging resource-constrained edge nodes such as Roadside Units (RSUs) as servers to deliver services to users \cite{MobAwareCache}, easing computational loads for users and enhancing their experience \cite{RSUBox}. One of the primary challenges in effective orchestration within the continuum is optimizing service placement, which involves selecting the most suitable functions for UE requests while jointly allocating computing and networking resources. This approach promotes resource sharing and maintains a deterministic system to ensure requests are met according to their latency requirements \cite{ACNC}. However, meeting stringent QoS requirements during high-velocity vehicular mobility remains challenging \cite{5GSurveyVehicle}. Also, mobility induces spatiotemporal heterogeneity in service demand, creating localized congestion where UE demands exceed edge node capacities. These challenges necessitate innovative strategies that address both the stochastic nature of vehicular traffic patterns and the limitations of conventional terrestrial infrastructure.

\IEEEpubidadjcol
\IEEEpubidadjcol

Unmanned Aerial Vehicles (UAVs) have significant potential to enhance the edge-cloud continuum's capabilities in service delivery. Conventional terrestrial networks, characterized by sparse distribution, often struggle to maintain consistent connections, especially on busy roads and during peak traffic hours. In this context, aerial-terrestrial networks, which are cost-effective and flexible, can be employed to provide prompt responses in demanding environments \cite{MultiUAVCoverage}. UAVs serve as aerial base stations and edge servers to deliver high-bandwidth services to ground-based UEs. Also, UAVs are considered integral components of the upcoming 6G landscape, playing a crucial role in the envisioned ubiquitous connectivity that supports bandwidth-intensive and real-time vehicular applications. However, as UAVs traverse diverse routes and engage with multiple UEs while managing a variety of computing requests, trajectory planning optimization becomes essential to ensure service coverage in UAV-assisted vehicular networks \cite{MECUAV, UAVTrajectoryDiffServ2024}. Furthermore, the variability of time-dependent channel dynamics presents a challenge for maintaining E2E latency in continuous service delivery, resulting from vehicles' high mobility \cite{MultiAgentDRL2022}. This necessitates the development of efficient trajectory and resource planning, as well as Multiple-Access Control (MAC) schemes, to mitigate mutual interference in a shared spectrum environment \cite{MACWireless, MACWCNC}.

Extant literature has advanced UAV-assisted vehicular networks; however, these approaches predominantly employ reactive mechanisms, artificially decouple the optimization of resource planning and MAC from trajectory planning, and insufficiently address composed service orchestration—collectively constraining system scalability and adaptability in dynamic environments. To fill in this gap, we propose a novel service orchestration framework for vehicular aerial-terrestrial 6G networks that integrates MAC, UAV trajectory optimization, and composed service placement within a heterogeneous edge-cloud continuum where both RSUs and UAVs function as communication interfaces for UEs. The main contributions and novelties of this paper are outlined as follows: 
\begin{itemize}
    \item A multi-time-scale Mixed Integer Non-Linear Programming (MINLP) formulation to optimize coverage and energy consumption of vehicular composed requests under E2E latency requirements in aerial-terrestrial networks.
    \item A decomposition of the problem into multi-UAV trajectory planning, MAC, and composed service placement for complexity reduction. To the best of our knowledge, this is the first work to consider these interconnected aspects while managing user interference on shared channels.
    \item A predictive Hierarchical Deep Reinforcement Learning (HDRL) framework that combines DRL with a Bayesian algorithm to enhance requests and channel quality prediction accuracy. The HDRL framework balances long-term and short-term objectives by facilitating interactions between the trajectory planning, MAC, and service placement modules, considering their distinct dynamics and time scales to achieve a globally optimal solution.
\end{itemize}

The upcoming sections of the paper are organized as follows. Section~\ref{sec:literature_review} provides a detailed overview of existing works and their limitations. Section~\ref{sec:system_model} details the system model, presenting the fundamental elements and their interactions. The problem formulation is introduced in Section~\ref{sec:problem}.
Section~\ref{sec:method} elaborates on the proposed method, with a detailed explanation of its design and implementation. Subsequently, Section~\ref{sec:results} presents the simulation settings, analyzes the convergence, and compares the proposed method's performance against baseline approaches. Finally, Section~\ref{sec:conclusion} encapsulates the study's key insights and future research directions.

\section{Related Works} \label{sec:literature_review}
The field of service orchestration in the vehicular edge-cloud continuum, vital for enabling next-generation vehicular applications, has witnessed advancements in recent years. Research in this domain addresses diverse dimensions such as the transition from single-UAV to multi-UAV scenarios, the evolution from heuristic approaches to adaptive and learning algorithms, and the shift from isolated challenges to complex, integrated problems, including joint trajectory planning and service provisioning \cite{sarkar2023artificial}. Table~\ref{tab_paper_comparison} provides a summary of the literature, offering a comparative analysis of existing works.

Given UAV mobility, the literature has focused on trajectory planning \cite{wei2023joint} or assuming deterministic, predefined mobility patterns \cite{fadlullah2020hcp}. There is increasing attention to trajectory planning solutions that leverage their agility for rapid deployment, reliable Line-of-Sight (LoS) connectivity, and the flexibility to adapt their coverage areas \cite{liu2019reinforcement}. In this regard, Santos \textit{et al.} \cite{santos2023mobility} proposed deploying multiple UAVs in underserved regions to ensure low-latency service delivery for mobile users. Similarly, Wei \textit{et al.} \cite{wei2023joint} tackled UAV trajectory planning while accounting for physical and environmental obstacles. 

Advancements in UAV mobility have spurred research on integrating joint trajectory planning and channel selection to optimize aerial-terrestrial networks with limited interfaces. As evidenced by Nabi \textit{et al.} \cite{nabi2024comprehensive}, which highlighted key challenges in aerial edge computing, including real-time adaptability and connectivity management for reliable communication. Due to this, some studies addressed these challenges by incorporating non-orthogonal multiple access in edge-cloud environments \cite{han2024drl}. Pervez \textit{et al.} \cite{powerTrajectory} proposed an iterative algorithm for user association, channel power allocation, and segment-based UAV trajectories in integrated aerial-terrestrial networks for smart vehicular services. Qin \textit{et al.} \cite{TrajectoryChannel} introduced a cluster-based air-ground integrated network with UAVs for access and high-altitude platforms for backhaul, \secH{optimizing UAV trajectories and subchannel selection to enhance energy and spectrum efficiency}. Further, other works focused on joint scheduling and channel selection in the edge-cloud environment, accounting for dynamic channel variations to minimize latency and energy consumption \cite{RASharedChannel}. Huang \textit{et al.} \cite{huang2024jointSAGINs} addressed the integration of satellite communications and aerial platforms through a DRL-based approach, optimizing both channel selection and trajectory planning. Qi \textit{et al.} \cite{qi2022energy} extended this line of research by proposing an energy-efficient framework combining content placement, spectrum allocation, co-channel pairing, and power control, improving channel selection and system performance.

\begin{table*}[!t]
\centering
\caption{Summary of existing schemes and comparison based on considerations in trajectory planning, MAC, and service placement.}
\vspace{-5pt}
\label{tab_paper_comparison}
\setlength{\tabcolsep}{4.5pt}
\footnotesize
\begin{tabular}{l c c c c c c c}
\toprule
\textbf{Reference} & 
\textbf{\secH{Objective Function}} & 
\textbf{Algorithm} & 
\textbf{\begin{tabular}[c]{@{}c@{}}Multi-\\UAV\end{tabular}} &
\textbf{\begin{tabular}[c]{@{}c@{}}\secH{UEs}\\\secH{Mobility}\end{tabular}} &
\textbf{\begin{tabular}[c]{@{}c@{}}Trajectory\\Planning\end{tabular}} &
\textbf{\begin{tabular}[c]{@{}c@{}}Multiple\\Access\end{tabular}} &
\textbf{\begin{tabular}[c]{@{}c@{}}Service\\Placement\end{tabular}} \\
\midrule

He \textit{et al.} \cite{he2024online} &
\secH{Maximize acceptance \!\&\! enhance energy} &
Actor-Critic \!\&\! Q-learning &
\checkmark &
\secH{Random} &
\checkmark &
-- &
\checkmark \\

SAC-TORA \cite{li2024service} &
\secH{Minimize energy consumption} &
Soft Actor-Critic &
\checkmark &
\secH{--} &
\checkmark &
-- &
\checkmark \\

Gupta \textit{et al.} \cite{gupta2023trajectory} &
\secH{Max-min aggregate throughput} &
SCA optimization &
\checkmark &
\secH{--} &
\checkmark &
-- &
-- \\

Qin \textit{et al.} \cite{qin2024joint} &
\secH{Minimize energy consumption} &
PMADDPG &
\checkmark &
\secH{--} &
\checkmark &
\checkmark &
\checkmark \\

Li \textit{et al.} \cite{GNNCruiseControl2023} &
\secH{Provisioning rate} &
GNN-DRL &
-- &
\secH{--} &
\checkmark &
-- &
-- \\

Muto \textit{et al.} \cite{UAVTrajectoryDiffServ2024} &
\secH{Minimize computational costs} &
Multi-agent DRL &
$\sim$ &
\secH{--} &
\checkmark &
-- &
\checkmark \\

\myHighlight{Dutriez \textit{et al.} \mbox{\cite{5GRelayElection}}} &
\secH{Maximize energy efficiency} &
\myHighlight{Deep Q-Network} &
\myHighlight{--} &
\secH{--} &
\myHighlight{--} &
\myHighlight{$\sim$} &
\myHighlight{--} \\

FL-SNTD3 \cite{li2024multi} &
\secH{Provisioning rate \& latency} &
Deep federated learning &
\checkmark &
\secH{--} &
\checkmark &
-- &
-- \\

DM-SAC-H \cite{huang2024jointSAGINs} &
\secH{Minimize energy consumption \& latency} &
Soft Actor-Critic &
\checkmark &
\secH{--} &
\checkmark &
-- &
\checkmark \\

HaDDQN \cite{qi2022energy} &
\secH{Energy efficiency} &
HaDDQN &
\checkmark &
\secH{Random} &
$\sim$ &
\checkmark &
\checkmark \\

Wei \textit{et al.} \cite{wei2023joint} &
\secH{Service execution success rate} &
Deep Q-Network &
-- &
\secH{--} &
\checkmark &
-- &
\checkmark \\

SCOFT \cite{chen2024joint} &
\secH{Minimize energy consumption} &
Hierarchical DRL (HDRL) &
\checkmark &
\secH{Random} &
\checkmark &
-- &
$\sim$ \\

\textbf{Proposed Solution} &
\secH{Maximize coverage \& optimize energy} &
Hierarchical DRL (HDRL) &
\checkmark &
\secH{Predictive} &
\checkmark &
\checkmark &
\checkmark \\

\bottomrule
\end{tabular}
\begin{tablenotes}
\footnotesize
\item SCA: Successive Convex Approximation; PMADDPG: Probabilistic Multi-Agent Deep Deterministic Policy Gradients.
\end{tablenotes}
\vspace{-8pt}
\end{table*}

Some studies have been carried out in the literature to study service provisioning and trajectory planning together for non-terrestrial networks. He \textit{et al.} \cite{he2024online} used an online DRL approach to investigate the interplay between continuous UAV trajectory planning and discrete service deployment actions. Le \textit{et al.} \cite{GNNCruiseControl2023} addressed optimization challenges in UAV-assisted edge networks, focusing on UAV trajectory and service provisioning in dynamic environments, using Graph Neural Networks (GNN) to optimize UAV speed, heading, and service deployment. Ning \textit{et al.} \cite{UAVTrajectoryDiffServ2024} developed a framework for UAV trajectory design that considers users' computational tasks and probabilistic service preferences. By facilitating decentralized trajectory optimization, they tried to minimize computational costs while maximizing service efficiency. 
Additionally, Li \textit{et al.} \cite{li2024service} explored a multi-UAV-enabled orchestration scheme for heterogeneous services, utilizing collaborative capabilities to minimize overall energy consumption in the system.


Despite innovative strategies, existing research on service orchestration in UAV-assisted networks still exhibits noteworthy limitations. Approaches assuming static users often fail to provide timely responses in highly dynamic environments \cite{gupta2023trajectory, qin2024joint, wei2023joint}. Several studies overlook the challenges of orchestrating composed services with diverse functions and shared resources across multiple users, which constrains scalability and flexibility \cite{GNNCruiseControl2023, 5GRelayElection, li2024multi, gupta2023trajectory}. Current solutions are predominantly reactive, adapting UAV trajectories and deployment only after receiving feedback, underscoring the need for proactive orchestration of future conditions like user mobility and demand.
\secH{For instance, SCOFT\mbox{\cite{chen2024joint}} optimized UAV trajectory and service placement for energy efficiency; however, its decisions remain reactive and are based solely on instantaneous system states, while we explicitly integrate predictive models of user mobility and service demand.}
While some works consider communication links between network elements\mbox{\cite{qi2022energy, qin2024joint}}, the complexities of \secH{dynamically sharing spectrum or considering channel qualities} for UAV-assisted service orchestration remain unexplored.
\secH{For example, HaDDQN does not predict channel dynamics and instead models them as stochastic processes, resulting in a reactive orchestration strategy that cannot anticipate future conditions. Likewise, Qin \textit{et al.}\mbox{\cite{TrajectoryChannel}} optimized decisions at a single control layer using only the current network state, without explicitly forecasting future demand.}
\secH{Recent advances in wireless techniques \mbox{\cite{11072365, 10882982}} improve spectral efficiency but are limited to communication-layer offloading and do not jointly address trajectory planning and composed service orchestration} Finally, the coupled optimization of trajectory planning, MAC, and service placement, each significantly influencing the others, has yet to be holistically addressed, which is essential for next-generation vehicular 6G service orchestration.

\section{System Model} \label{sec:system_model}
\begin{figure}[t!]\centering
\includegraphics[width=3.4in]{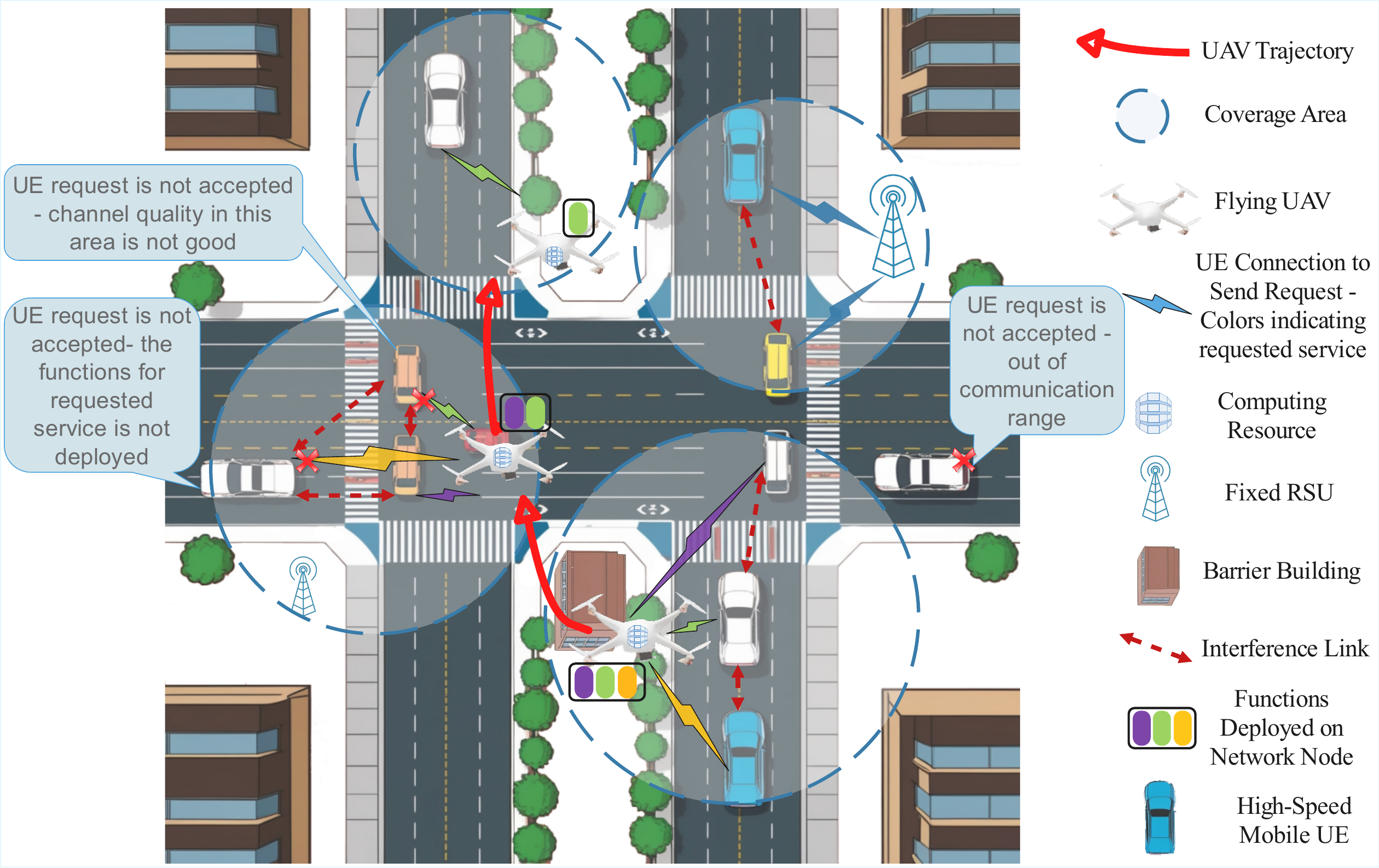}
\vspace{-8pt}
  \caption{System Model: Supporting UEs through RSUs and UAVs with service coverage, quality channel access, and deployed function availability.}
    \vspace{-4pt}
    \label{figure:systen_model_intro}
\end{figure}

This section describes the system's detailed structures: network architecture, services, and interactions between the UEs, RSUs, and UAVs across the network, as shown in Fig.~\ref{figure:systen_model_intro}.

\vspace{-8pt}

\subsection{Vehicular Network Architecture}
In this paper, a tiered aerial-terrestrial network is investigated, denoted by $\mathcal{G}(\mathcal{N}, \mathcal{L}, \mathcal{P})$, which encompasses the coexistence of vehicular networks and the edge-cloud continuum. The network integrates $\boldsymbol{\mathbbm{N}}$ computing nodes, such as core nodes, RSUs, and UAVs, each equipped with networking capabilities. They connected through links to deliver services over designated areas during uniform time frames, indexed by $t$. Nodes close to UEs provide limited computation capabilities at a high cost, whereas core nodes possess significant computing power with lower resource expenses \cite{farhoudi2024}. Each node $n$ is characterized by its processing capability $\widehat{\mathcal{C}}_n$
and its energy budget $\overline{\mathcal{E}}_n$.
Wired and wireless links connect the different network elements, the set of which is denoted by $\boldsymbol{\mathbbm{L}}^t \subset \{ l : (n, n\myprime)|n, n\myprime \in \boldsymbol{\mathbbm{N}} \}$\footnote{Mobility of both UEs and UAVs leads to a dynamically changing network topology, and UEs should stay within the coverage area of an RSU or UAV to maintain a connection; otherwise, links between them become unavailable.}. Each link $l$ is associated with a bandwidth capacity $\widehat{\mathcal{L}}_l$
and transmission energy consumption $\overline{\xi}_l$.
Packets associated with request $r$ that traverse link $l$ experience latencies at time frame $t$ determined using a function denoted as $\mathcal{D}^{t}_{r,l}$, which is deterministically computed as a function of the current network state, link length, and link load. 
Based on the links available at each time frame, a set of paths, $\boldsymbol{\mathbbm{P}}^t$, is available for packet transmissions. Nodes forming a particular path $p$ are represented as $\mathbbm{N}_p$, where $\mathcal{H}^t_p$ and $\mathcal{T}^t_p$ are head and tail nodes respectively, with $\mathcal{J}^t_{p,l}$ denoting the inclusion of link $l$ in path $p$ in time fame $t$.

\vspace{-5pt}
\subsection{Services}
Services are characterized by outlining essential aspects, including their functions, data model, and data graph \cite{FarhoudiComposition}. Each composed service $\!s \!\in\! \boldsymbol{\mathbbm{S}}\!\!\!=\!\!\!\{ 1, 2, ..., \mathcal{S} \}\!$ is segmented in functions, denoted as $\!\boldsymbol{\mathbbm{F}}_s\!=\!\!\{ 1, 2, ..., \mathcal{F}_{s} \}$, where each function implemented by virtual instances. The data model accounts for the complex interdependencies between these functions, ensuring that service execution starts with the initial function and proceeds with sequential or parallel execution of subsequent functions. The data graph, $\mathcal{G}_{s}$, outlines the structure of the composed service, including inputs, outputs, preconditions, and results, along with the total service duration time $\overrightarrow{\mathcal{T}_s}$. 

\vspace{-5pt}
\subsection{Vehicular User Equipment}
The set of vehicular UEs is defined as \(\boldsymbol{\mathbbm{U}} = \{ 1, 2, \ldots, \mathcal{U} \}\). UEs generate requests at various time intervals for vehicle-to-everything communication, enabling them to transmit requests via appropriate wireless communication protocols in heterogeneous networks. The first node a UE connects to is known as the Point of Attachment (PoA), through which UE requests are handled to reach desired services. Requests $r \! \in \! \{1,\dots,\mathcal{R}\}$ are sent by UEs, and $u_r$ identifies who generates the request $r$. Each request $r$ enters the system at time $\mathcal{T}'_r$, requesting a composed service $\mathcal{S}_r$ over a predefined duration, which means the request should be processed within $\Delta_r = [\mathcal{T}'_r, \mathcal{T}'_r + \overrightarrow{\mathcal{T}_s}]$. In each time frame, the number of active requests may vary due to factors like mobility patterns and bandwidth-saving strategies. Requests come with specific requirements, including network bandwidth $\widecheck{\mathcal{L}}_{r}$, each atomic function's minimum capacity $\widecheck{\mathcal{I}}_{r,f}$, and E2E latency $\widecheck{\mathcal{D}}_{r}$ \cite{globecom2023}. Successful service delivery entails meeting the capacity and QoS requirements.

\vspace{-5pt}
\subsection{Network Areas}
The network is divided into areas denoted by $\boldsymbol{\mathbbm{A}}\!\!=\!\!\{ 1, 2, ..., \mathcal{A} \}$ to ensure comprehensive coverage. The dimensions of areas vary based on geographical factors like obstacles.
\secH{The assumption is adopted for analytical and computational tractability, where UAV movement is represented at the area level rather than a fully continuous flight trajectory, within which UAVs reposition or hover within areas to serve users. This abstraction is known to capture the dominant mobility effects with negligible loss of accuracy at the considered time scale\mbox{\cite{UAVDiscretization}}.}
UEs exhibit dynamic behavior by frequently moving between areas, while being assumed to remain within a single area during each time frame to simplify movement modeling. The area of UEs who send a request $u_r$ at time frame $t$ is called $\mathcal{A}^{t}_{u_r,a}$. Computing nodes are located in various parts of the network, with core nodes and RSUs placed in fixed areas and UAVs moving in different areas. 


The energy consumption of UAVs traveling between the areas $a_1$ and $a_2$ is given by $\Lambda(a_1,a_2)$ \myHighlight{{\eqref{eq_uav_energy}}, which captures hovering and propelling energy\mbox{\cite{omarUAVUAG}}. The hovering power includes $a_1$ and $a_2$ travel duration $\Delta_{a_1,a_2}$, induced power coefficient $I$, UAV's total weight $W_n$, air density $\varphi$, and rotor disk area $\upsilon_r$. The movement power accounts for aerodynamic drag, where $\varsigma$ and $\upsilon_f$ represent the drag coefficient and frontal area, and $V_w(t)$ is the UAV's instantaneous velocity.} 

\footnotesize
\vspace{-6pt}
\begin{align}
\label{eq_uav_energy}
\Lambda(a_1,a_2)
= \int_{0}^{\Delta_{a_1,a_2}}
\Bigg(
\frac{I \cdot W_n^{3/2}}{\sqrt{2 \cdot \varphi \cdot \upsilon_r}}
+ \frac{1}{2} \cdot \varsigma \cdot \varphi \cdot \upsilon_f \cdot V_w(t)^3
\Bigg)\, dt
\end{align}
\vspace{-8pt}
\normalsize

\subsection{Wireless Channel Model}
The wireless channel model is defined with a focus on uplink transmissions, wherein UE requests are dynamically scheduled to minimize the collision probability in each time frame through effective channel assignment. Each time frame $t$ is subdivided into $\mathfrak{T}_t$ smaller time slots to enable fine-grained MAC. The set of available channels is denoted by $\boldsymbol{\mathbbm{C}}\!\!=\!\!\{ 1, 2, ..., \mathcal{C} \}$, and each channel $c$ is associated with a specific energy consumption $\overline{\mathcal{M}}_c$ and channel quality $\widecheck{\mathcal{Q}}^{\tau}_{c, a}$ in area $a$, as energy requirements vary with \myHighlight{operating frequency and path loss.} To mitigate collisions resulting from simultaneous transmissions over the same channel within different locations, our model accommodates multiple uplink channels and allows channel reuse in neighboring areas. Meanwhile, we assume that the downlink channels used for service responses to UEs are collision-free, ensuring reliable response transmission.

UE and UAV movements lead to time-varying channel conditions, where the absence of a LoS link can degrade E2E transmission quality. To model this, $\theta_{a}^{\text{\tiny{LoS}}}$ is defined, which represents the likelihood of establishing a LoS connection in area $a$, determined by environmental density, UAV altitude, and weather conditions \cite{liu2019reinforcement, disaster6G}. When a LoS connection exists, transmission quality is assumed to be ideal; otherwise, under Non-LoS (NLoS) conditions occurring with probability $1-\theta_{a}^{\text{\tiny{LoS}}}$, signal quality is governed by an instantaneous channel gain ${H}_{c, a}^{\tau}$ that incorporates path loss, small-scale fading, and weather-dependent absorption $\Omega_{\tau}$, expressed as {\eqref{channl_instantaneous}}. 
In this expression, $R_{c,a}^{\tau}$ and $\chi_{c,a}^{\tau}$ denote the Rayleigh and lognormal components, $D_0$ is the reference distance and $d_{c,a}^{\tau}$ denotes the transmitter-receiver separation on channel $c$ in area $a$ at time slot $t$. Also, the parameters $(\nu_s,\eta_s(\Omega_{\tau}))$ correspond to the path loss exponent and weather-dependent shadowing factor under LoS/NLoS propagation, and $\zeta(\Omega_{\tau})$ models atmospheric attenuation such as fog or rain. The received signal-to-noise ratio (SNR) for the channel is given by {\eqref{channl_quality}}, where $P_{tx}$ represents the transmit power, $\mathcal{N}_0$ is the thermal noise spectral density, and $\Delta f$ is the subcarrier spacing. 
Following real-world 6G multiple-access modeling \mbox{\cite{5GRelayElection}}, a transmission is considered successful if $\gamma_{c,a}^{\tau}$ exceeds a predefined quality threshold $\widehat{\mathcal{Q}}$ (determined by QoS requirements). Accordingly, the resulting binary channel-quality indicator $\mathcal{Q}_{c,a}^{\tau}$ equals 1 for successful transmissions (either through LoS or when NLoS \myHighlight{SNR} is acceptable) and 0 when signal degradation becomes excessive\footnote{Intuitively, $\mathbbm{1}(C)$ equals one if the condition $C$ is satisfied.}, thereby capturing the essential dynamics of UAV communication reliability \myHighlight{under variable weather conditions.}



\begingroup
\footnotesize
\setlength{\abovedisplayskip}{0pt}
\setlength{\abovedisplayshortskip}{0pt}
\setlength{\belowdisplayskip}{0pt}
\setlength{\belowdisplayshortskip}{0pt}
\begin{align}
\label{channl_instantaneous}
H_{c,a}^{\tau} = R_{c,a}^{\tau} \cdot 10^{\chi_{c,a}^{\tau} \cdot \eta_s(\Omega_{\tau})/10}
\cdot (D_0 / d_{c,a}^{\tau})^{\nu_s}
\cdot 10^{-\zeta(\Omega_{\tau})/10}
\end{align}
\endgroup

\footnotesize
\begin{align}
\label{channl_quality}
\widecheck{\mathcal{Q}}^{\tau}_{c,a}
=
\begin{cases}
1, & \mbox{\small{w.p.}} \ \theta_{a}^{\text{\tiny{LoS}}}, \\[2pt]
\mathbbm{1}\!\Big( \gamma_{c,a}^{\tau} =
\frac{P_{tx} \cdot H_{c,a}^{\tau}}{\mathcal{N}_0 \cdot \Delta f}
\geq \widehat{\mathcal{Q}}
\Big),
& \mbox{\small{w.p.}} \ 1-\theta_{a}^{\text{\tiny{LoS}}}
\end{cases}
\end{align}
\normalsize



\section{Problem Formulation} \label{sec:problem}
In this section, the formulation of a MINLP optimization problem, termed energy-Aware muLtipLe-access service OrChestration for vehicular Aerial-TErrestrial networks (ALLOCATE) is presented. The problem pertains to optimizing energy-efficient service coverage through the placement of requested service functions on network nodes, the allocation of a specific set of functions to each request based on its service graph, the assignment of a channel and path for each request to facilitate data delivery from its PoA to the corresponding functions, and the subsequent return of data to the originating PoA. Table~\ref{table:notations} shows the notations used in ALLOCATE.

\vspace{-5pt}
\subsection{Objective Function}
The objective function is formulated to maximize request acceptance - ensuring comprehensive service coverage - while minimizing energy consumption (OF). A scaling factor ($\alpha$) is introduced to balance the trade-off between energy consumption and request acceptance, adjusting the relative importance of the two factors to ensure an optimal solution. \myHighlight{It not only aligns with theoretical optimization goals but also reflects the practical constraints and behaviors of 6G aerial-terrestrial networks \mbox{\cite{disaster6G}}}. The total energy consumption $\overline{\mathcal{W}}$ (C1) accounts for the prioritization of nodes, from edge to cloud, with varying computational capabilities, communication channels, and links, as well as the energy required for UAVs to traverse between candidate areas \myHighlight{(incorporates propulsion dynamics derived from the aerodynamic energy model (Eq.~{\eqref{eq_uav_energy}}))}. For a request to be satisfied, all required functions during the request's duration should be deployed (C2).

\renewcommand{\arraystretch}{1.15}
\begin{table}[t!]
\centering
\caption{List of notations used in the problem formulation.}
\vspace{-6pt}
\label{table:notations}
\setlength{\tabcolsep}{3pt}
\footnotesize
\begin{tabular}{cl}
\toprule
\textbf{Notation} & \textbf{Description} \\
\midrule
$\mathcal{G}(\mathcal{N}, \mathcal{L}, \mathcal{P})$ & Vehicular aerial-terrestrial edge-cloud network \\
$\mathcal{G}_{s}$ & Service $s$ data graph \\

$t \in \boldsymbol{\mathbbm{T}}$ & Time frame (of Total service time) \\
$\tau \in {\mathfrak{T}_t}$ & Time slot (of time frame $t$) \\
$\overrightarrow{\mathcal{T}_s}$ & Total time for delivering service $s$ \\
$\mathcal{T}'_r$ & Entry time of UE request $r$ \\
$\widecheck{\mathpzc{T}}_r$ & Minimum required time slots to send request $r$ \\


$\boldsymbol{\mathbbm{N}} / \boldsymbol{\mathbbm{A}}$ & Set of network nodes / predefined areas \\
$\boldsymbol{\mathbbm{L}}^t / \boldsymbol{\mathbbm{P}}^t$ & Set of (wireless links / active paths) at time $t$ \\
$\boldsymbol{\mathbbm{U}} / \boldsymbol{\mathbbm{R}} / \boldsymbol{\mathbbm{C}}$ & Set of (active UEs / Requests / uplink channels) \\

$\widehat{\mathcal{C}}_n / \overline{\mathcal{E}}_n$ & (Processing / Energy consumption) of node $n$ \\
$\widehat{\mathcal{L}}_l / \overline{\xi}_l$ & (Bandwidth capacity / Transmission energy) of link $l$ \\

$\mathcal{H}^t_p / \mathcal{T}^t_p$ & Head/Tail node of path $p$ at time frame $t$ \\
$\mathcal{J}^t_{p,l}$ & The inclusion of link $l$ in path $p$ at time frame $t$ \\
$\mathcal{D}^{t}_{r,l}$ & Latency experienced by request $r$ over link $l$ at time $t$ \\

$\mathcal{F}_f \in \boldsymbol{\mathbbm{F}}_s$ & Atomic function $f$ (of functions set) \\

$u_{r} / \mathcal{S}_r$ & (UE who send / Composed service) of request $r$ \\
$\mathcal{A}^{t}_{u,a}$ & Indicator for UE $u$ located in area $a$ at time $t$ \\
$\overline{\mathcal{M}}_c$ & Energy consumption for using uplink channel $c$ \\
$\widecheck{\mathcal{Q}}^{\tau}_{c,a}$ & Quality of uplink channel $c$ in area $a$ at time slot $\tau$ \\
$\widecheck{\mathcal{L}}_{r}$ & Network bandwidth required for request $r$ \\
$\widecheck{\mathcal{I}}_{r,f}$ & Minimum capacity required for function $f$ of request $r$ \\
$\widecheck{\mathcal{D}}_{r}$ & Latency requirement for request $r$ \\

$\tilde{\mathcal{X}}^{t}_{r,f}$ & if function $f$ of request $r$ is selected at time frame $t$ \\
$\tilde{\mathcal{Y}}^{t}_{f,n}$ & if function $f$ is placed on node $n$ at time frame $t$ \\
$\tilde{\mathcal{Z}}^{\tau}_{r,c}$ & if channel $c$ is selected for request $r$ at time slot $\tau$ \\
$\tilde{\mathcal{S}}^{t}_{n,a}$ & if node $n$ is deployed in area $a$ at time frame $t$ \\
$\tilde{\mathcal{B}}^{t}_{u,n}$ & if UE $u$ is connected (binned) to node $n$ at time frame $t$ \\
$\overrightarrow{\mathcal{R}}^{t}_{r,p}$ & if path $p$ is selected for request $r$ at time frame $t$ \\
\bottomrule
\end{tabular}
\vspace{-16pt}
\end{table}

The selection of key components within the system is governed by binary decision variables: $\tilde{\mathcal{X}}^{t}_{r,f}$ indicates whether request $r$ is served by function $f$, $\tilde{\mathcal{Y}}^{t}_{f,n}$ denotes the hosting node $n$ for the function $f$, and $\tilde{\mathcal{Z}}^{\tau}_{r,c}$ represents the selection of channel $c$ for request $r$. Also, $\tilde{\mathcal{S}}^t_{n, a}$ specifies the candidate area $a$ for network node $n$ while RSU areas are always fixed, $\tilde{\mathcal{B}}^t_{u,n}$ represents the PoA of UE $u$ at time frame $t$ while they are moving, and $\overrightarrow{\mathcal{R}}^{t}_{r,p}$ determines if path $p$ is selected to send request $r$ packets to deployed nodes and receive the response. 

\footnotesize
\begin{align*}\label{OBJECTIVEFUNCTION1}
     & \text{ALLOCATE: } \mathrm{ max } \sum_{\boldsymbol{\mathbbm{R}}}\left(\tilde{\mathcal{X}}_r \right) - \alpha \cdot \overline{\mathcal{W}} \qquad \textit{ s.t.} \quad \text{C1 - C12.} \tag{OF} 
    \\
    & \overline{\mathcal{W}} \triangleq \sum_{\boldsymbol{\mathbbm{F}}_{s}, \boldsymbol{\mathbbm{N}}, \boldsymbol{\mathbbm{T}}}
    \tilde{\mathcal{Y}}^{t}_{f,n} \cdot \overline{\mathcal{E}}_n + \sum_{\boldsymbol{\mathbbm{N}}, \boldsymbol{\mathbbm{A}}, \boldsymbol{\mathbbm{T}}} \Lambda( \tilde{\mathcal{S}}^{t+1}_{n,a_1}-\tilde{\mathcal{S}}^{t}_{n,a_2})  \\
    & + \sum_{\boldsymbol{\mathbbm{L}}^t, \boldsymbol{\mathbbm{P}}^t, \boldsymbol{\mathbbm{R}}, \Delta_r}   
        \overline{\xi}_l \cdot \mathcal{J}^t_{p,l} \cdot \overrightarrow{\mathcal{R}}^{t}_{r,p} +
    \sum_{\boldsymbol{\mathbbm{R}},\boldsymbol{\mathbbm{C}}, \boldsymbol{\mathbbm{T}}, \mathfrak{T}_t}\tilde{\mathcal{Z}}^{\tau}_{r,c} \cdot \overline{\mathcal{M}}_c \tag{C1} 
    \\
    &
    \tilde{\mathcal{X}}_r = \prod_{\boldsymbol{\mathbbm{F}}_{s_r}, \Delta_r} \tilde{\mathcal{X}}^t_{r,f} \qquad \qquad \qquad \qquad \qquad \qquad \quad \forall r \in \boldsymbol{\mathbbm{R}} \tag{C2}
\end{align*}
\normalsize

\vspace{-5pt}
\subsection{Constraints}
Constraints ensure: (1) appropriate channel allocation within network areas (MAC protocol); (2) optimal service deployment on network nodes with efficient packet routing from PoAs to service nodes (service placement); and (3) dynamic UAV adjustment to accommodate active service requests (trajectory planning). All optimization processes simultaneously satisfy capacity limitations and QoS requirements. Notably, the system operates on multi time-scales: time frames (denoted by $t$), and time slots (denoted by $\tau$) with each time frame comprising ${\mathfrak{T}_t}$ time slots. All resource allocation tasks, except for channel selection, operate on time frame granularity, while channel selection functions at a higher frequency of time slots.

\textbf{Channel Selection:} 
Efficient service delivery in a multiple-access environment necessitates avoiding simultaneous transmissions over the same channel in an area to prevent collisions. For each UE, no more than one channel should be selected for transmitting its request (C3). Other UEs within the same area should be prevented from using the same channel at the same time slot (C4). This is vital due to the limited available channels of sufficient quality ($\widecheck{\mathcal{Q}}^{\tau}_{c, a}$)\myHighlight{, derived from the wireless channel model for request transmission Eq.~{\eqref{channl_quality}}}. It indicates UAVs' inability to handle multiple channels simultaneously to maintain real-world consistency between link admission and channel quality. Requests are sent through the channels of nodes to which UEs are directly connected, affecting energy consumption $\overline{\mathcal{W}}$ with the selected transmission channel.

\begingroup
\footnotesize
\setlength{\abovedisplayskip}{-4pt}
\setlength{\abovedisplayshortskip}{-4pt}
\begin{align*}\label{channel_selection}
     & \sum_{\boldsymbol{\mathbbm{C}}} \tilde{\mathcal{Z}}_{r, c}^{\tau} \leq 1 \qquad \qquad \qquad \qquad \qquad \qquad \forall r, \tau \in \boldsymbol{\mathbbm{R}}, \bigcup_{t \in \Delta_r}^{}{\mathfrak{T}_t} \tag{C3} 
    \\
    & \sum_{\boldsymbol{\mathbbm{R}}}
    \tilde{\mathcal{Z}}_{r, c}^{\tau} \cdot \mathcal{A}^t_{u_r,a} \leq 1 \qquad \qquad \qquad \qquad \forall c,a, \tau \in \boldsymbol{\mathbbm{C}}, \boldsymbol{\mathbbm{A}}, \bigcup_{t \in \boldsymbol{\mathbbm{T}}}^{}{\mathfrak{T}_t} \tag{C4} 
\end{align*}
\endgroup


\textbf{Function Placement:}
When dealing with composed services, it is necessary to consider the deployment of various functions of services. Thus, each function targeted by at least one request should be deployed on an available network node for the duration of the request (C5). Moreover, each request $r$ should be assigned to appropriate functions based on its required service $\boldsymbol{\mathbbm{F}}_{s_r}$, only if they are transmitted on a quality channel not used by other UEs during (C6). \myHighlight{This constraint, along with C4, assesses whether a UE's transmissions over quality channels} meet the minimum required time slots ($\widecheck{\mathpzc{T}}_r$). If they do, the service could be provided; otherwise, the variable $\tilde{\mathcal{X}}^{t}_{r,f}$ is set to zero (request is not accepted).

\vspace{-5pt}
\footnotesize
\begin{align*}\label{service_placement}
     & \sum_{\boldsymbol{\mathbbm{N}}} \tilde{\mathcal{Y}}^{t}_{f,n} \geq \left( \sum_{\boldsymbol{\mathbbm{R}}} \tilde{\mathcal{X}}^{t}_{r,f} \right) / \mathcal{R} \qquad \qquad \quad \quad \quad \forall f, t \in \boldsymbol{\mathbbm{F}}_s, \boldsymbol{\mathbbm{T}} \tag{C5}
    \\ \\
    & \tilde{\mathcal{X}}^{t}_{r,f} \leq (\!\!\!\sum_{\boldsymbol{\mathbbm{C}}, \boldsymbol{\mathbbm{A}}, \mathfrak{T}_t}\!\!{\tilde{\mathcal{Z}}_{r, c}^{\tau} \cdot \widecheck{\mathcal{Q}}^{\tau}_{c,a} \cdot \mathcal{A}^{t}_{u_r,a}} ) / \widecheck{\mathpzc{T}}_r  \qquad \forall r,f,t \in \boldsymbol{\mathbbm{R}}, \boldsymbol{\mathbbm{F}}_{s_r}, \Delta_r \tag{C6}
\end{align*}
\normalsize

\textbf{Path Selection:}
For the effective transmission of inquiry traffic from a UE to its designated nodes, where the requested service is deployed, and the subsequent return of the response, feasible E2E routes should be provided. A unique inquiry path $\overrightarrow{\mathcal{R}}^{t}_{r,p}$ is established for each request, originating at the UE's PoA ($\tilde{\mathcal{B}}^t_{u,n}$). Considering UE mobility, the response will be directed to the PoA corresponding to the location where the UE will be present when the request duration concludes, thereby addressing its mobility. The requested service's functions are interconnected based on $\mathcal{G}_{s_r}$ to reach their final destination, with the last function sending the response to $u_r$ (C7).


\footnotesize
\vspace{-6pt}
\begin{align}
\label{path_selection}
\!\!\!\!\!\!\!\!\sum_{\substack{
p \in \boldsymbol{\mathbbm{P}}^t,\;
\mathbbm{N}_p,\;
\boldsymbol{\mathbbm{F}}_{s_r} \\
\mathcal{H}^t_p = \sum_{\boldsymbol{\mathbbm{N}}} n \cdot \tilde{\mathcal{B}}^{\mathcal{T}'_r}_{u_r,n} \\
\mathcal{T}^t_p = \sum_{\boldsymbol{\mathbbm{N}}} n \cdot \tilde{\mathcal{B}}^{\mathcal{T}'_r+\overrightarrow{\mathcal{T}_{s_r}}}_{u_r,n}
}} \!\!\!\!\!\!\!\!\!\!\!\!\!
\overrightarrow{\mathcal{R}}^{t}_{r,p}
\cdot \mathbbm{1}\!\left( \tilde{\mathcal{Y}}^{t}_{f,n} = 1 \right)
= 1
\qquad \qquad \quad \forall r,\, t \in \boldsymbol{\mathbbm{R}},\, \Delta_r
\tag{C7}
\end{align}
\vspace{-4pt}
\normalsize

\textbf{Capacity:} 
Given the network's limited capabilities, it is essential to manage nodes' and links' capacities to maintain system stability. So, the total number of requests allocated to any node does not exceed its processing capacity (C8). Additionally, each link's capacity should not be exceeded during request and response transmission (C9). 

\vspace{-5pt}
\footnotesize
\begin{align*}\label{capacity_constraints}
    &\sum_{\boldsymbol{\mathbbm{R}}, \boldsymbol{\mathbbm{F}}_{s_r} }\tilde{\mathcal{X}}^{t}_{r,f} \cdot \tilde{\mathcal{Y}}^{t}_{f,n} \cdot  \widecheck{\mathcal{I}}_{r,f} \leq \widehat{\mathcal{C}}_n \qquad \qquad \qquad \qquad \quad \forall n, t \in \boldsymbol{\mathbbm{N}}, \boldsymbol{\mathbbm{T}} \tag{C8} 
    \\
    &\sum_{\boldsymbol{\mathbbm{R}}, \boldsymbol{\mathbbm{P}}^t} \mathcal{J}^t_{p,l} \cdot \overrightarrow{\mathcal{R}}^{t}_{r,p} \cdot \widecheck{\mathcal{L}}_{r} \leq \widehat{\mathcal{L}}_{l} \qquad \qquad \qquad \qquad \quad \quad \forall l, t \in \boldsymbol{\mathbbm{L}}^t, \boldsymbol{\mathbbm{T}} \tag{C9}
\end{align*}
\vspace{-2pt}
\normalsize

\textbf{UAV Trajectory Planning:}
Trajectory planning is essential for covering UEs and meeting latency requirements while minimizing energy consumption. As the objective is to optimize energy consumption and changes in UAV locations affect energy usage in $\overline{\mathcal{W}}$, the optimization problem prompts to limit UAV mobility while ensuring vehicular UE connectivity. Each network node should remain within exactly one area during each time frame (C10) and each UE should be associated with one PoA within its area at a specified time frame (C11).

\vspace{-5pt}
\footnotesize
\begin{align*}\label{trajectory_planning}
    & \sum_{\boldsymbol{\mathbbm{A}}} \tilde{\mathcal{S}}^{t}_{n,a} 
    = 1
    \qquad \qquad \qquad \qquad \qquad \qquad \qquad \quad \forall n, t \in \boldsymbol{\mathbbm{N}}, \boldsymbol{\mathbbm{T}} \tag{C10} \\
     & \sum_{\boldsymbol{\mathbbm{N}},\boldsymbol{\mathbbm{A}}}\tilde{\mathcal{B}}^{t}_{u_r,n} \cdot \tilde{\mathcal{S}}^{t}_{n,a} \cdot \mathcal{A}^{t}_{u_r,a} \leq 1
    \qquad \qquad \qquad \qquad \forall r,t \in \boldsymbol{\mathbbm{R}}, \Delta_r \tag{C11}
\end{align*}
\normalsize

\textbf{QoS Requirements:}
Ensuring timely and consistent service delivery is paramount to meeting UEs' stringent QoS expectations. Constraint (C12) sets a maximum acceptable latency for request handling, verifying that if a request is accepted, its latency requirements are met. This constraint prevents UEs from monopolizing acceptance based solely on low energy consumption, ensuring requests' QoS within the specified requirements. \myHighlight{The latency threshold $\widecheck{\mathcal{D}}_{r}$ aggregates transmission, propagation, and processing delays from $\mathcal{D}^{t}_{r,l}$ to capture realistic E2E latency. Coupled with the channel-quality indicator $\widecheck{\mathcal{Q}}^{\tau}_{c,a}$, only transmissions meeting the required SNR threshold are accepted, ensuring compliance with latency and reliability standards in UAV-assisted vehicular networks \mbox{\cite{3gpptR38901}}.}

\vspace{-6pt}
\footnotesize
\begin{align*}\label{qos_requirements}
    & 
    \sum_{\boldsymbol{\mathbbm{P}^t}, \boldsymbol{\mathbbm{L}}^t, \Delta_r} \mathcal{J}^t_{p,l} \cdot \mathcal{D}^{t}_{r,l} \cdot \overrightarrow{\mathcal{R}}^{t}_{r,p}
    \leq \widecheck{\mathcal{D}}_{r} \qquad \qquad \qquad \qquad \quad \forall r \in \boldsymbol{\mathbbm{R}} \tag{C12}
\end{align*}
\normalsize

\subsection{Complexity Analysis}\label{subsec:complexity}
The ALLOCATE problem is classified as NP-hard, reducible from the multidimensional knapsack problem presented in \cite{faticanti_cutting_2018}. This classification implies a worst-case computational complexity proportional to the solution space size \cite{pataki_basis_2010}. Determining the optimal solution for a set of requests requires interdependent evaluations: analyzing each UAV node in every area (trajectory planning), assessing each channel in each time slot (channel selection), and considering every node, function, and path (placement). As any allocation for any request in a given time frame impacts and is influenced by allocations made for other requests, complexity arises. Consequently, all permutations of UAVs, requests, and times should be examined, creating an exponentially large solution space $\boldsymbol{\mathbbm{T}}!(\boldsymbol{\mathbbm{N}}_{\text{\tiny{UAV}}}! \boldsymbol{\mathbbm{A}}) \cdot (\boldsymbol{\mathbbm{C}} \mathfrak{T}_t) \cdot (\boldsymbol{\mathbbm{N}} \boldsymbol{\mathbbm{F}}_s \boldsymbol{\mathbbm{P}} \boldsymbol{\mathbbm{U}}!)$.

In addition to the inherent complexity, in dynamic networks characterized by UAV and UE mobility, several key parameters remain uncertain. Without prior knowledge of UE areas ($\mathcal{A}^{t}_{u,a}$) and their request arrivals, it is impossible to determine the appropriate channels to send requests ($\tilde{\mathcal{Z}}^{\tau}_{r,c}$) or assess channel qualities ($\widecheck{\mathcal{Q}}^{\tau}_{c,a}$). Consequently, UAV trajectory planning ($\tilde{\mathcal{S}}^{t}_{n,a}$) is impossible ahead of time, as the UE's locations dictate UAV movement. The uncertainty also extends to function placement ($\tilde{\mathcal{Y}}^{t}_{f,n}$) and path planning ($\overrightarrow{\mathcal{R}}^{t}_{r,p}$), as the links connecting UAVs and network nodes are not predetermined. Therefore, tackling ALLOCATE requires a novel method that accommodates the dynamic nature and inherent uncertainties in the network.

\begin{figure}[t!]\centering
\includegraphics[width=3.2in]{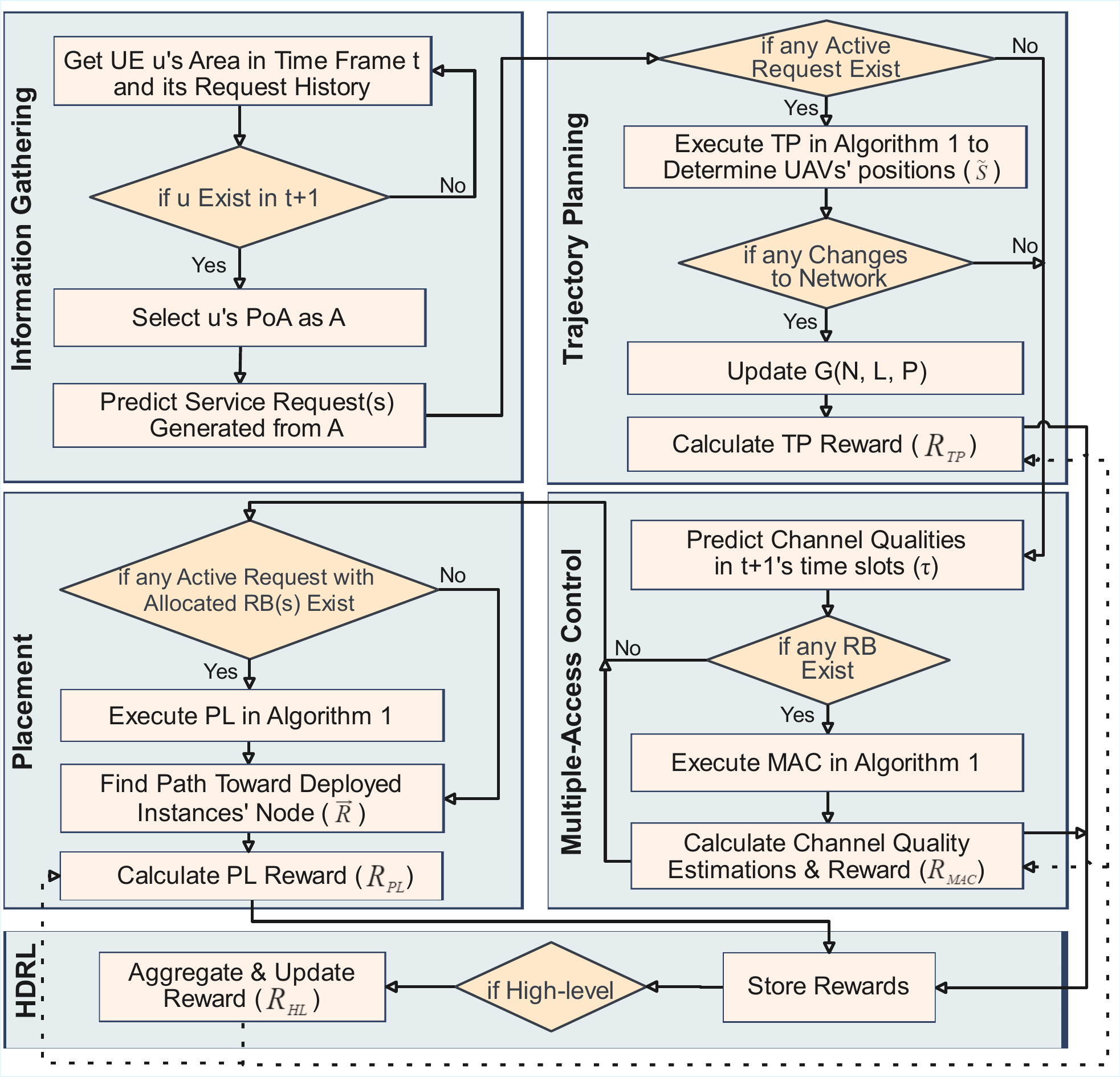}
\vspace{-10pt}
  \caption{The proposed method's process for time frame $t$, including Information Gathering and Service Orchestration phases.}
    \label{figure:flowchart}
\end{figure}

\section{Proposed method} \label{sec:method}
We propose \underline{p}r\underline{e}dictive ene\underline{r}gy e\underline{f}ficient s\underline{e}rvice or\underline{c}hes\underline{t}ration (PERFECT) to tackle ALLOCATE, which operates in two phases: Information Gathering and Service Orchestration. \myHighlight{As illustrated in Fig.~{\ref{figure:flowchart}}, the PERFECT workflow is presented as a flowchart depicting the sequential execution and interconnection of its key components. The Information Gathering phase employs predictive learning to capture the temporal evolution of UE mobility and service request dynamics. Specifically, the process begins with the collection of request histories and network states, followed by the generation of predicted mobility and service demands.} Using information retrieved from the former, the latter allocates resources. To further manage the complexity of large-scale problems, the Service Orchestration phase is divided into three sub-problems: Trajectory Planning (TP), MAC, and Placement (PL) modules. \myHighlight{The TP module determines UAV trajectories and PoA updates based on predicted UE distributions; its output defines feasible communication links and coverage areas for the next time slot. The MAC module subsequently manages channel access to mitigate interference and updates the channel quality estimations for the next time frame. Finally, the PL module decides where network functions should be deployed by evaluating node capacities, latency constraints, and energy efficiency. The outputs of these modules are looped back into the environment, updating the state for the next orchestration cycle.}

\begin{figure}[t!]\centering
\includegraphics[width=3in]{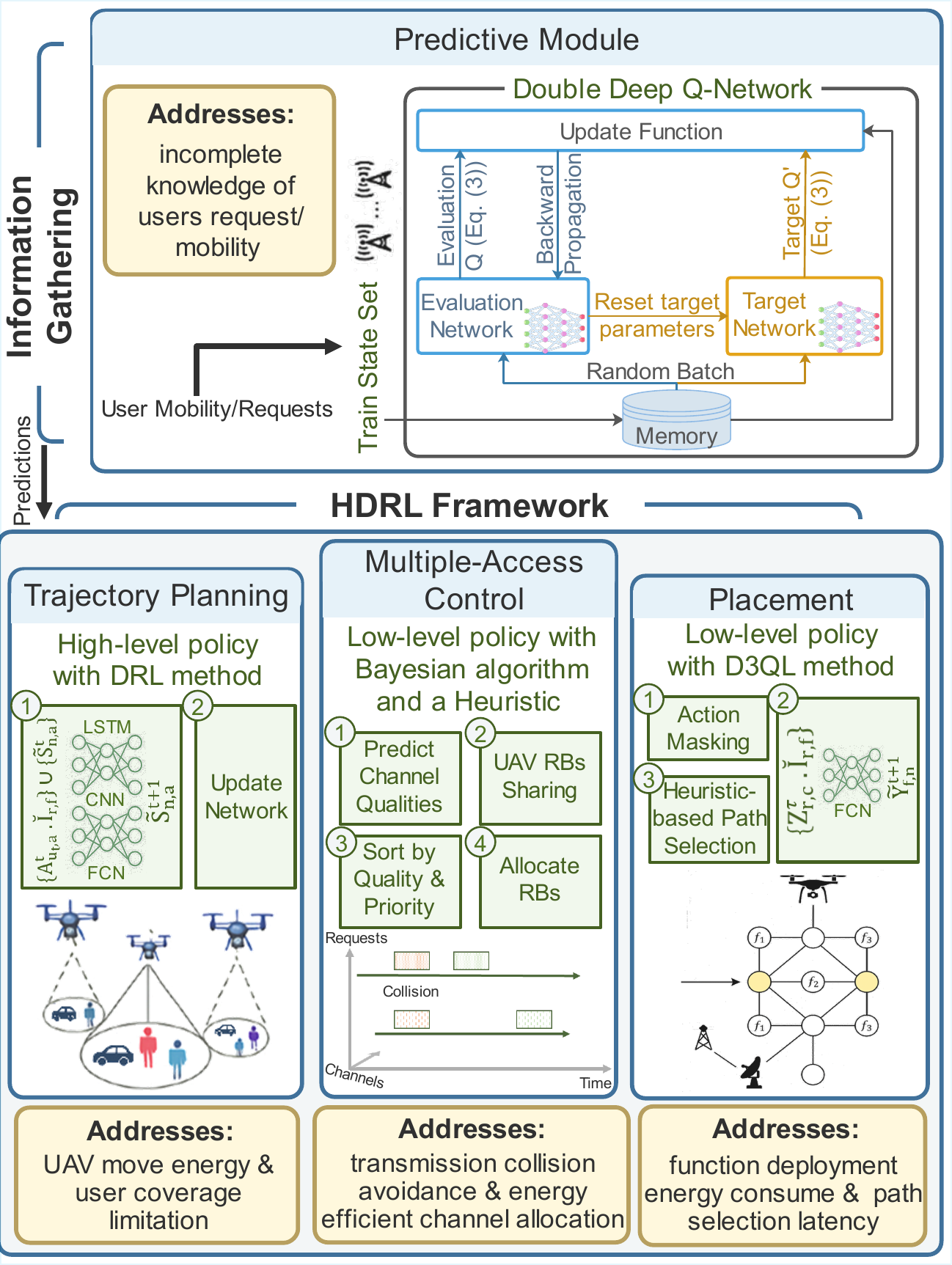}
\vspace{-6pt}
  \caption{\myHighlight{PERFECT's technologies and addressed challenges. The predictive module forecasts UE mobility and request patterns, while the HDRL framework integrates TP, MAC, and PL modules.}}
    \label{figure:technologies}
    \vspace{-4pt}
\end{figure}

\myHighlight{Fig.~{\ref{figure:technologies}} outlines a complementary perspective by detailing technical underpinnings and addressing challenges. The Information Gathering phase, implemented via a Double Deep Q-Network (DDQN), resolves imperfect knowledge limitations. The DRL-based TP ensures energy-efficient UAV movement while maximizing coverage; the MAC's heuristic channel allocation achieves collision-aware transmission by evaluating channel qualities; and the action masking-enhanced Dueling Double Deep Q-Learning (D3QL) PL enables energy-efficient, latency-optimized function deployment and path selection.}

\vspace{-5pt}
\subsection{Motivation for Learning Algorithms}
To facilitate adaptive decision-making in dynamic and uncertain scenarios, DRL demonstrates promise. \myHighlight{Classical value-based approaches like} DQL, a DRL's potential techniques, \myHighlight{approximate action-value functions through deep Neural Networks (NNs)}. Double DQL \myHighlight{enhances the stability of} DQL by decoupling action selection and evaluation, utilizing the target value defined in \eqref{eq_DDQL_target} \cite{hasselt_deep_2016}. This target incorporates reward $R$, state $O$, real action $a$, and predicted action $a'$ to update DQL weights ($\mathcal{W}$) for each observation-action of time $t$ ($O^t, a^t$). Here, $a'\!\!\!=\!\! \text{argmax}_{a \in \boldsymbol{\mathcal{A}}} Q({O}^{t+1}, a; \mathcal{W}^t)$ with $\mathcal{W}$ representing evaluation weights updated at each step, and $\mathcal{W}^-$ as target weights, synchronized every $\hat{t}\!\gg\!0$ step. D3QL enhances DDQL by integrating Wang \textit{et al.}'s dueling approach \cite{wang2016dueling}. In D3QL, separate estimators calculate state values ($\mathcal{V}$) and action advantages ($\rho$), combining them to compute Q-values \eqref{eq_dueling} and weights \eqref{eq_update_w}. This approach improves training stability, accelerates convergence, and mitigates overestimation issues. 

\footnotesize
\begin{equation}
\begin{aligned}
     Y^t  = {R}^{t} + \Gamma \cdot Q({O}^{t+1}, a'; {\mathcal{W}}^{t-})
\end{aligned}
\label{eq_DDQL_target}
\end{equation}
\vspace{-3mm}
\normalsize


\footnotesize
\vspace{-18pt}
\begin{align}
\label{eq_dueling}
Q(O^t, \!a^t; \!\mathcal{W}^t)
\!=\! \mathcal{V}(O^t;\! \mathcal{W}^t)
\!+\! \rho(O^t,\! a^t; \mathcal{W}^t)
\!-\! \frac{1}{\left| \boldsymbol{\mathcal{A}} \right|}
\!\!\sum_{a^{\prime} \in \boldsymbol{\mathcal{A}}}\!\!
\rho(O^t,\! a^{\prime};\! \mathcal{W}^t)
\end{align}
\normalsize

\vspace{-10pt}
\footnotesize
\begin{equation}
\begin{aligned}
     \mathcal{W}^{t+1} \gets \mathcal{W}^{t} + \sigma \cdot [Y^t - Q(O^t, a^t; \mathcal{W}^t)] \nabla_{\mathcal{W}^t} Q(O^t, a^t; \mathcal{W}^t)
\end{aligned}
\label{eq_update_w}
\end{equation}
\vspace{-3mm}
\normalsize

\myHighlight{Although effective in small and moderate-sized decision spaces, these non-hierarchical methods operate on a single action space and uniform time scale. Consequently, when applied to heterogeneous processes with multi-node orchestration, they exhibit degraded convergence, become sensitive to state dimensionality, and require large exploration budgets. Also, policy-gradient methods such as Proximal Policy Optimization (PPO) alleviate some instability issues through clipped surrogate objectives, yet they also suffer from slow convergence under combinatorial action spaces and lack mechanisms for decomposing multi-time-scale decisions.}

\myHighlight{To handle these limitations in large state and action spaces, HDRL employs action-space and temporal abstraction, decomposing orchestration into coordinated high- and low-level modules. This design enhances training efficiency by enabling module interaction across distinct dynamics and timescales: high-level policies manage long-term strategy, while low-level policies handle short-term control. Each module operates within a tailored state-action domain, reducing learning complexity compared to monolithic DQN/PPO models. Hierarchical coupling allows high-level policies to integrate low-level outcomes, improving stability, sample efficiency, and scalability with increasing UAVs, UEs, and functions. To achieve a globally optimal solution and resolve the limited observation capability, careful coordination to harmonize the modules' behaviors is needed. By designing well-structured state representations and employing action masking in low-level policies, the proposed HDRL architecture avoids convergence issues common in flat DRL. Thus, PERFECT achieves faster convergence, high service quality, and superior performance under multi-timescale constraints, establishing HDRL as an effective solution for complex large-scale problems.}

\footnotesize
\begin{algorithm}[t!]
\label{alg_resource_allocation}
\caption{Service Orchestration Phase}
\KwInput{$\boldsymbol{\mathbbm{T}}$, $\epsilon'$, and $\widetilde{\epsilon}$}
\KwOutput{$\tilde{\mathcal{S}}$, $\tilde{\mathcal{B}}$, $\tilde{\mathcal{Z}}$, $\tilde{\mathcal{Y}}$, $\overrightarrow{\mathcal{R}}$}
$\mathcal{W}_\text{\tiny{TP}, \tiny{PL}} \leftarrow \mathbf{0}$, ${\mathcal{W}^-}_\text{\tiny{TP}, \tiny{PL}} \leftarrow \mathbf{0}$, $\epsilon_\text{\tiny{TP}, \tiny{PL}} \gets 1$, $\psi_\text{\tiny{TP}, \tiny{PL}} \gets \{\}$\\
\For{$t$ in $[1:\boldsymbol{\mathbbm{T}}]$}
{
    \textcolor{gray}{$\star$ Trajectory Planning \myHighlight{(high-level, frame-scale)} $\star$}  \\
    $\tilde{\mathcal{S}}^{t+1} \gets \textit{EpsilonGreedy}\Big(Q({O}_\text{\tiny{TP}}^{t}, {A}_\text{\tiny{TP}}^{t}; \mathcal{W}_{\text{\tiny{TP}}}), \epsilon_{\text{\tiny{TP}}} \Big)$ \\
    $\mathcal{H} \gets \{ t-\mathcal{H}, \ldots, t \}$ \\
    Calculate ${O}_\text{\tiny{TP}}^{t+1}$ according to $\mathcal{H}$ and \eqref{trajectory_planning_state} \\
    $\epsilon_{\text{\tiny{TP}}} \gets \max(\epsilon_{\text{\tiny{TP}}} - \epsilon', \widetilde{\epsilon} ) $ \\
    Update $\mathcal{G}(\mathcal{N}, \mathcal{L}^t, \mathcal{P}^t)$ based on $\tilde{\mathcal{S}}^{t+1}$ \\
    Select $\tilde{\mathcal{B}}^{t+1}$ for each request \\
    
    \textcolor{gray}{$\star$ \myHighlight{Multiple-Access Control (low-level, slot-scale)} $\star$}  \\
    Calculate ${O}_\text{\tiny{MAC}}^{t+1}$ according to \eqref{mac_state} \\
    Calculate $\overline{\mathcal{Q}}^{t+1}$ based on ${O}_\text{\tiny{MAC}}^{t+1}$ \eqref{mac_action} \\
    $\tilde{\mathcal{Z}} \gets $ \textit{MAC}\Big($\tilde{\mathcal{B}}^{t+1}, \tilde{\mathcal{S}}^{t+1}, \omega^{t}, \overline{\mathcal{Q}}^{t+1}, \mathfrak{T}_t $ \Big) \\
    \myHighlight{Compute ${R}_\text{\tiny{MAC}}^{t}$ based on channel qualities} \\

    \textcolor{gray}{$\star$ Placement \myHighlight{(low-level, frame-scale)} $\star$}  \\
    $\tilde{\mathcal{Y}}^{t+1} \gets \textit{EpsilonGreedy}\Big(Q({O}_\text{\tiny{PL}}^{t}, {A}_\text{\tiny{PL}}^{t}; \mathcal{W}_{\text{\tiny{PL}}}), \epsilon_{\text{\tiny{PL}}} \Big)$ \\
    Calculate ${O}_{\text{\tiny{PL}}}^{t+1}$ according to \eqref{placement_state} \\
    $\epsilon_{\text{\tiny{PL}}} \gets \max(\epsilon_{\text{\tiny{PL}}} - \epsilon', \widetilde{\epsilon} ) $ \\
    Select $\overrightarrow{\mathcal{R}}^{t+1}$ for each request \\
    
    \textcolor{gray}{$\star$ Training and Reward Propagation $\star$}  \\
    Calculate ${R}_\text{\tiny{TP}}^{t}$ and ${R}_\text{\tiny{PL}}^{t}$ based on \eqref{trajectory_planning_reward} and \eqref{placement_reward} \\
    \If{high level}{
        ${R}_\text{\tiny{HL}}^{t} = {R}_\text{\tiny{TP}}^{t} + \chi \cdot {R}_\text{\tiny{MAC}}^{t} + \kappa \cdot {R}_\text{\tiny{PL}}^{t}$ \\
        ${R}_\text{\tiny{TP}}^{t} , {R}_\text{\tiny{PL}}^{t} = {R}_\text{\tiny{HL}}^{t}$ \\
        \myHighlight{Update global state to sync TP $\rightarrow$ MAC/PL}
    }

    $\psi_\text{\tiny{TP}} \gets \psi_\text{\tiny{TP}} \cup \{({O}_\text{\tiny{TP}}^{t}, \tilde{\mathcal{S}}^{t+1}, {R}_\text{\tiny{TP}}^{t}, {O}_\text{\tiny{TP}}^{t+1})\}$ \\
    Train $\mathcal{W}_{\text{\tiny{TP}}}$ on batch of samples from $\psi_\text{\tiny{TP}}$ \eqref{eq_update_w}  \\
    
    $\psi_\text{\tiny{PL}} \gets \psi_\text{\tiny{PL}} \cup \{({O}_\text{\tiny{PL}}^{t}, \tilde{\mathcal{Y}}^{t+1}, {R}_\text{\tiny{PL}}^{t}, {O}_\text{\tiny{PL}}^{t+1})\}$ \\
    Train $\mathcal{W}_{\text{\tiny{PL}}}$ on batch of samples from $\psi_\text{\tiny{PL}}$ \eqref{eq_update_w} \\
}
\end{algorithm}
\normalsize

\vspace{-7pt}
\subsection{Information Gathering Phase}

The Information Gathering phase constructs a dynamic network graph $\mathcal{G}(\mathcal{N}, \mathcal{L}, \mathcal{P})$, representing UE areas and anticipated requests for the next time frame. Following the strategy by Farhoudi \textit{et al.} \cite{globecom2023}, we adopt an online model to handle requests' sporadic and dynamic nature, as offline machine learning models cannot adapt to rapid changes in request patterns and render them insufficient for accurate predictions. 

\myHighlight{The proposed framework inherently addresses real-time demand fluctuations and unpredictable mobility patterns through its learning-based approach.} \myHighlight{Specifically,} each PoA is equipped with a D3QL agent, \myHighlight{continuously updating its policy based on newly observed transitions, enabling rapid adaptation to sudden traffic or request changes.}
These agents predict the probability of each request $r$ being issued in the next time frame, identifying the presence of $u_r$ in the area $a$ ($\mathcal{A}^{t+1}_{u_r,a}$). During the prediction process, the agent returns a prioritized list of requests with the highest likelihood as well, with a reward based on the prediction accuracy and the state consisting of the received requests' history and the previous UE areas.
\myHighlight{Unlike static predictors, our online DRL method exploits} Short-Term Memory (LSTM) layers \myHighlight{to capture temporal variations in request arrivals caused by high UE mobility, while} Convolutional Neural Network (CNN) layers \myHighlight{extract spatial correlations between adjacent areas.} A memory bank stores observed transitions to enable efficient NN updates through random sampling\myHighlight{, as illustrated in Fig.{~\ref{figure:technologies}}}. 

\begin{figure*}[t!]\centering
\includegraphics[width=6.7in]{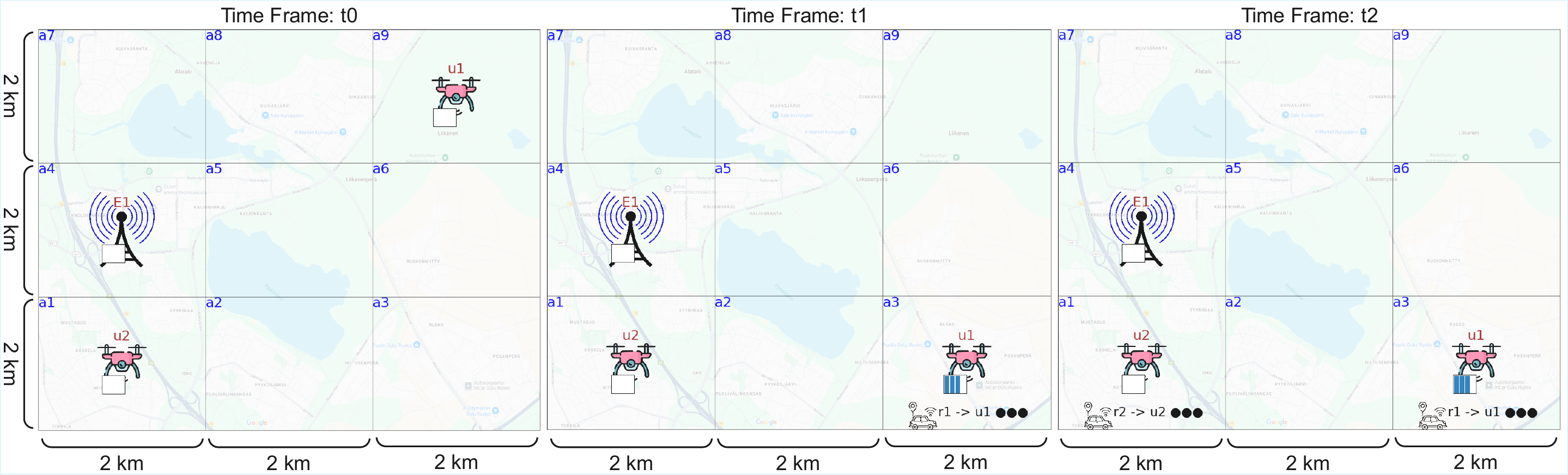}
\vspace{-9pt}
  \caption{A possible scenario where UE $r_1$ enters from area $a_3$ in time frame $t_1$. Although $u_2$ requires less energy to reach $a_3$ than $u_1$, TP learning algorithm decides to move $u_1$, predicting that $r_2$ will enter from $a_1$. \myHighlight{Each grid cell in this deployment across Oulu city represents approximately 2km {×} 2km.}}
    \vspace{-9pt}
    \label{figure:total_movement}
\end{figure*}

\vspace{-6pt}
\subsection{Service Orchestration Phase}\label{serviceOrchestrationPhase}
Following centralized aggregation of UE area predictions and anticipated service requests for the subsequent time frame, this phase deploys functions and allocates resources to meet predicted requests and requirements. To tackle the multi-faceted challenges of decision-making in dynamic, resource-constrained environments, we employ an HDRL framework. It facilitates decision-making by decomposing complex problems into manageable subproblems, employing high-level policies for strategic, long-term decisions and low-level policies for operational, short-term ones. 
\myHighlight{The hierarchical interaction perfectly fits the proposed decomposition approach and ensures immediate action adaptation to real-time network dynamics without deviating from long-term objectives. Reward-sharing and feedback mechanisms enable continuous synchronization of short-term adaptations with global optimization goals.}

\footnotesize
\begin{algorithm}[t!]\label{epsilon_greedy_algorithm}
\caption{Learning's \textit{EpsilonGreedy} Process}
\KwInput{$Q(S,A;W)$, $\epsilon$}
\KwOutput{Actions}
$\zeta \gets$ randomly generate a number from $[0:1]$ \\
\If{$\zeta > \epsilon$}
{
    $Actions \gets$ argmax$_{a \in A} Q({O}, a; \mathcal{W})$ \\
}
\Else
{
    $Actions \gets$ select random actions
}
\end{algorithm}
\normalsize

The TP module serves as the high-level policy, responsible for \myHighlight{long-term} UAV movement \myHighlight{decisions that optimize service coverage, feasible communication links, and the structural conditions under which low-level modules operate}. The MAC and PL modules operate as low-level policies, \myHighlight{responsible for real-time, fine-grained decisions; such as channel allocation, Resource Block (RB) management, function deployment, and path selection--based on rapidly changing environmental conditions.} Each RB represents a time slot within a specific channel, determining the allocation of communication resources for transmitting UE requests \myHighlight{and enabling the MAC to react at a higher temporal resolution than the TP}. Using a simultaneous learning approach (Algorithm~\ref{alg_resource_allocation}), \myHighlight{the low-level policies are first stabilized independently to ensure reliable short-term behavior}. MAC learns efficient channel allocation across time slots, while PL optimizes function deployment and path selection. \myHighlight{Their decisions shape the environment observed by TP, and their instantaneous rewards form the components of the high-level reward.} Once \myHighlight{stabilized, their operational outcomes (aggregated rewards) are integrated into the training of high-level policies, forming a closed feedback loop where updated short-term results influence long-term optimization.} Hence, ${R}_\text{\tiny{HL}}^{t}$ is considered the total reward, which combines rewards from TP, MAC, and PL modules, scaled by factors $\chi$ and $\kappa$ to balance their contributions. Through this \myHighlight{dynamic interaction across distinct time scales, low-level policies optimize within TP-defined constraints while TP learns to anticipate downstream effects, maintaining coherence between times and achieving real-time adaptability without sacrificing efficiency.}

\textbf{Trajectory Planning:}
The TP module determines optimal UAV areas for the upcoming time frame, aiming to optimize total UAV movement energy consumption while maximizing request coverage. The purpose of using a learning algorithm rather than a heuristic approach lies in its ability to achieve long-term optimization, thereby reducing UAVs' overall energy consumption during their movement. Using a D3QL algorithm, detailed in Algorithm~\ref{alg_resource_allocation} (steps 5–10), it predicts UAV movements based on past observations, prioritizing long-term energy optimization. A scenario is illustrated in Fig.~\ref{figure:total_movement} \myHighlight{that represents the Oulu city area, where each grid cell represents approximately 2km {×} 2km}. In this scenario, a UE request $r_1$ arises in area $a_3$ in time frame $t_1$, with UAVs $u_1$ existing at $a_9$ and $u_2$ at $a_1$ in $t_0$. Although $u_1$ requires more energy to reach $a_3$ (based on $\Lambda$), moving $u_1$ is more efficient, predicting another request at $a_1$ and avoiding unnecessary movements for $u_2$. In another scenario, if $r_1$ moves from $a_3$ at $t_1$ to $a_2$ at $t_2$, the algorithm pre-positions $u_3$ at $a_2$ from $t_0$, minimizing movement energy while ensuring timely delivery.

The module implementation approach is to design the state and action spaces in a scalable manner. The state ${O}_\text{\tiny{TP}}^{t}$ incorporates total requested capacities per area and UAV locations, encoded in the one-hot format \eqref{trajectory_planning_state}, based on the last $\mathcal{H}$ observations. This representation is request number independent, making it scalable for networks with varying UE numbers. The state serves as input to a NN architecture consisting of LSTM, CNNs, and linear layers. The action ${A}_{\text{\tiny{TP}}}^{t}$ is then generated that represents areas assigned to UAVs for the next time frame, categorized as $\tilde{\mathcal{S}}^{t+1}$ \eqref{trajectory_planning_action}. The action results in changes to the network graph, as UAVs would be relocated across different areas, leading to alterations in the links and paths. Hereafter, $\tilde{\mathcal{B}}^{t+1}$ (PoAs) are determined based on $\mathcal{A}^{t+1}_{u,a}$ (UE areas) and $\tilde{\mathcal{S}}^{t+1}$ (network node areas). The reward function ${R}_\text{\tiny{TP}}^{t}$ maximizes coverage while optimizing UAV energy consumption, aligning with ALLOCATE's objective \eqref{trajectory_planning_reward}. This approach balances UE demand, service coverage, and energy efficiency, using predictive insights.

\vspace{-6pt}
\footnotesize
\begin{equation}
\begin{aligned} 
     {o}_\text{\tiny{TP}}^{t} &= \Big\{ \sum_{\boldsymbol{\mathbbm{R}}, \boldsymbol{\mathbbm{F}}_{s_r} } \mathcal{A}^t_{u_r,a} \cdot \widecheck{\mathcal{I}}_{r,f} \ | a \in \ \boldsymbol{\mathbbm{A}} \Big\} \cup \Big\{ \tilde{\mathcal{S}}^t_{n,a} | n, a \in  \boldsymbol{\mathbbm{N}}, \boldsymbol{\mathbbm{A}} \Big\}  \\
     {O}_\text{\tiny{TP}}^{t} &= \Big\{ {o}_{\text{\tiny{TP}}}^{h} | h \in \{ t-\mathcal{H}, \ldots, t \} \Big\}
\end{aligned}
\label{trajectory_planning_state}
\end{equation}
\vspace{-10pt}
\normalsize

\vspace{-10pt}
\footnotesize
\begin{equation}
\begin{aligned} 
{A}_{\text{\tiny{TP}}}^{t} &= \Big\{ \tilde{\mathcal{S}}^{t+1}_{n,a} | n, a \in  \boldsymbol{\mathbbm{N}}, \boldsymbol{\mathbbm{A}} \Big\}
\end{aligned} 
\label{trajectory_planning_action}
\end{equation}
\vspace{-10pt}
\normalsize

\vspace{-10pt}
\footnotesize
\begin{equation}
\begin{aligned} 
     {R}_\text{\tiny{TP}}^{t} &= \sum_{\boldsymbol{\mathbbm{R}}, \boldsymbol{\mathbbm{N}}}{\tilde{\mathcal{B}}^{t+1}_{u_r,n}} - \alpha \cdot \sum_{\boldsymbol{\mathbbm{N}}, \boldsymbol{\mathbbm{A}}} \Lambda( \tilde{\mathcal{S}}^{t+1}_{n,a_1}-\tilde{\mathcal{S}}^{t}_{n,a_2})
\end{aligned}
\label{trajectory_planning_reward}
\end{equation}
\vspace{-7pt}
\normalsize

\begin{table*}[t]
\centering
\caption{\myHighlight{Mapping of PERFECT Subproblems to system roles, decision layers, and applied Technologies.}}
\vspace{-8pt}
\label{table:subproblems_mapping}
\begin{tabular}{
>{\centering\arraybackslash}m{1.9cm}
>{\centering\arraybackslash}m{4.9cm}
>{\centering\arraybackslash}m{2.5cm}
>{\centering\arraybackslash}m{7.2cm}}
\toprule
\textbf{Subproblem} &
\textbf{Physical Aspect} &
\textbf{Decision Layer} &
\textbf{Applied Technology (Rationale)} \\
\midrule

\textbf{Info Gathering} &
UE mobility \& demand prediction &
Sensing \& Prediction &
D3QL \!\!+\!\! CNN \!\!+\!\! LSTM (spatiotemporal feature extraction) \\

\textbf{TP} &
UAV mobility, \!coverage \!\&\! energy \!efficiency &
Mobility \!management  &
D3QL (long-term control, stable convergence) \\

\textbf{MAC} &
Channel allocation \& latency control &
Communication/MAC &
Bayesian \!+\! heuristic (probabilistic–deterministic) \\

\textbf{PL} &
Function deploy, routing, \& QoS assure &
Resource allocation &
D3QL \!+\! masking \!+\! heuristic route (energy-efficient placement) \\

\textbf{HDRL} &
Cross-layer coordination across multi-time-scale &
Hierarchical control &
Policy integration (temporal abstraction, scalability) \\
\bottomrule
\end{tabular}
\vspace{-10pt}
\end{table*}

\textbf{Multiple-Access Control:}
The MAC module allocates energy-efficient channels ($\overline{\mathcal{M}}_c$) to UEs for request transmission. It functions at the time slot level, with each time frame containing $\mathfrak{T}_t$ time slots. Efficient allocation requires predicting channel qualities ($\widecheck{\mathcal{Q}}^{\tau}_{c,a}$) in each area and mapping them to requests. Thus, the module is divided into two parts: channel quality prediction and heuristic-based channel assignment. Since channel predictions should update dynamically during HDRL training periods, this component is included in the MAC module rather than the Information Gathering phase. This hybrid design prioritizes predictive elements for critical tasks while employing deterministic algorithms for simpler processes, ensuring energy-efficient channel allocation.

To predict channel quality, the MAC module utilizes a Bayesian algorithm that models and updates beliefs under uncertainty. The Bayesian algorithm assigns an initial quality value of $0.5$ to each channel in every area. As UAVs traverse across areas, these estimates (posterior beliefs) are revised incrementally based on updated observations ${O}_\text{\tiny{MAC}}^{t}$ \eqref{mac_state} and weighted by $\lambda$, which governs observation impact. This process, outlined in steps 12–13 of Algorithm~\ref{alg_resource_allocation}, continuously refines channel quality estimates $\overline{\mathcal{Q}}_{c, a}^{t}$, calculated using \eqref{mac_action}.

\footnotesize
\begin{equation}
\begin{aligned}
     {O}_\text{\tiny{MAC}}^{t} &= \Big\{ \frac{1}{\left| \mathfrak{T}_t \right|} \cdot \sum_{\mathfrak{T}_t}{ \widecheck{\mathcal{Q}}^{\tau}_{c,a} } | c, a \in  \boldsymbol{\mathbbm{C}}, \boldsymbol{\mathbbm{A}} \Big\}
\end{aligned}
\label{mac_state}
\end{equation}
\vspace{-6pt}
\normalsize

\vspace{-6pt}
\footnotesize
\begin{equation}
\begin{aligned}
     \overline{\mathcal{Q}}^t_{c,a} &= \Big\{ \lambda \cdot {O}_\text{\tiny{MAC}}^{t} + (1-\lambda) \cdot \overline{\mathcal{Q}}^{t-1}_{c,a}  \Big\}
\end{aligned}
\label{mac_action}
\end{equation}
\vspace{-6pt}
\normalsize

Following channel quality prediction, the MAC module allocates them to requests, as detailed in Algorithm~\ref{alg_MAC}. The allocation strategy prioritizes proximity to request deadlines to meet E2E latency requirements $\widecheck{\mathcal{D}}_{r}$. At the beginning of each time frame, co-located UAVs share available RBs. UAVs then allocate their RBs to connected UEs using predicted qualities $\overline{\mathcal{Q}}^{t}_{c, a}$, priorities $\omega^t$, and required time slots $\widecheck{\mathpzc{T}}_r$. Channels and requests are sorted by their respective quality and priority, and RBs are assigned iteratively until either RBs are exhausted or request requirements are fulfilled. Specifically, the algorithm calculates the minimum between remaining and required time slots ($\acute{\tau}$) for each request. It allocates all slots within the interval $\tau:\acute{\tau}$ if no prior allocation exists, adhering to the single-channel transmission constraint (C3). This procedure ensures high-priority UEs access superior-quality channels, optimizing resource usage while complying with QoS constraints.

\footnotesize
\begin{algorithm}[t!]\label{alg_MAC}
\caption{Channel Allocation Method}
\KwInput{$\tilde{\mathcal{B}}^{t+1}, \tilde{\mathcal{S}}^{t+1}, \omega^{t}, \overline{\mathcal{Q}}^{t+1}, \mathfrak{T}_t $}
\KwOutput{\{$\tilde{\mathcal{Z}}^{\tau} | \tau \in \mathfrak{T}_t \}$}
$\tilde{\mathcal{Z}}_{r, c}^{\tau} \gets 0 \quad \forall \ r, c, \tau \in \boldsymbol{\mathbbm{R}}, \boldsymbol{\mathbbm{C}}, \mathfrak{T}_t$ \\
\ForEach{$n$ in $\boldsymbol{\mathbbm{N}}$}
{
$\boldsymbol{\mathbbm{C}}_{sorted} \gets \textit{SORT}_{c} \big\{ \boldsymbol{\mathbbm{C}}, \textit{key}= \overline{\mathcal{Q}}_{c,a}^{t+1} | \tilde{\mathcal{S}}^{t+1}_{n,a} = 1 \big\}$ \\
$\boldsymbol{\mathbbm{R}}_{sorted} \gets \textit{SORT}_{r} \big\{ \boldsymbol{\mathbbm{R}}, \textit{key}= {\omega^t} | \tilde{\mathcal{B}}_{u_r,n}^{t+1} = 1 \big\}$ \\
\ForEach{$c$ in $\boldsymbol{\mathbbm{C}}_{sorted}$}
{
    $\tau \gets 0$ \\
    \While{$\tau \leq {\left| \mathfrak{T}_t \right|}$}
    {
        \ForEach{$r$ in $\boldsymbol{\mathbbm{R}}_{sorted}$}
        {
            $\acute{\tau} \gets \tau + min(\widecheck{\mathpzc{T}}_r, {\left| \mathfrak{T}_t \right|} - \tau)$ \\
            \If{$\sum_{c, \tau}{\tilde{\mathcal{Z}}_{r, c}^{ {\tau: \acute{\tau}} } == 0}$}
            {
                $\tilde{\mathcal{Z}}_{r, c}^{ {\tau: \acute{\tau}}} \gets 1$ \\
                $\tau \gets \acute{\tau}$ \\
            }
        }
    }
}
}
\end{algorithm}
\normalsize

\textbf{Placement:}
This module determines the optimal deployment of functions on network nodes $\tilde{\mathcal{Y}}^{t}_{f,n}$ and the paths to reach them $\overrightarrow{\mathcal{R}}^{t}_{r,p}$, relying on predicted requests' UE areas and calculated UAV locations. It aims at minimizing energy consumption for function deployment ($\overline{\mathcal{E}}_n$) while adhering to node processing and link bandwidth constraints ($\widehat{\mathcal{C}}_n$, $\widehat{\mathcal{L}}_l$), reducing transmission energy ($\overline{\xi}_l$), and meeting latency requirements ($\widecheck{\mathcal{D}}_{r}$). Given the impact of UE and UAV mobility on placement dynamics, the PL module uses a D3QL learning algorithm (Algorithm~\ref{alg_resource_allocation}, steps 16–19) to strategically deploy functions on nodes, ensuring that the total energy consumption is efficient and request requirements are in compliance over time. 

PL module implementation provides a highly efficient and deployable state space, designed for scalability and independence from request volume. Its state ${O}_{\text{\tiny{PL}}}^{t}$ includes total requested function capacities across all nodes with channel access as well as nodes' available capacities and associated energy consumption \eqref{placement_state}. Additionally, the reward function ${R}_\text{\tiny{PL}}^{t}$ focuses on maximizing accepted requests (coverage) while optimizing energy consumption, considering latency constraints \eqref{placement_reward}. Furthermore, directly considering all network nodes, functions, links, and paths in the action space poses scalability and convergence challenges due to the large action space. Two key strategies are employed to overcome this challenge. First, path selection is decoupled from learning via a heuristic integrated into reward calculation. Specifically, after selecting nodes for function deployment, the algorithm prioritizes requests by latency requirements and identifies feasible paths that meet the requirements with minimal energy consumption ($\overline{\xi}_l$). The algorithm penalizes invalid deployments with negative rewards, while otherwise rewarding the total energy consumed to establish the connection. Second, action masking reduces the effective action space without sacrificing optimality by eliminating infeasible actions by assigning large negative values. The infeasible actions include deploying functions with no predicted requests or exceeding thresholds. The PL module action ${A}_{\text{\tiny{PL}}}^{t}$ is defined as in \eqref{placement_action}.

\vspace{-8pt}
\footnotesize
\begin{equation}
\begin{aligned}\label{placement_state}
     {O}_{PL}^{t} &= \Big\{ \sum_{\boldsymbol{\mathbbm{R}}, \boldsymbol{\mathbbm{C}}, \mathfrak{T}_t} \tilde{\mathcal{Z}}_{r, c}^{\tau} \cdot \widecheck{\mathcal{I}}_{r,f} \ | f \in \ \boldsymbol{\mathbbm{F}}_{s} \Big\} \cup \Big\{ \big( \widehat{\mathcal{C}}_n, \overline{\mathcal{E}}_n \big) | n \in  \boldsymbol{\mathbbm{N}} \Big\}
\end{aligned}
\end{equation}
\vspace{-8pt}
\normalsize

\vspace{-8pt}
\footnotesize
\begin{equation}
\begin{aligned} 
&{A}_{\text{\tiny{PL}}}^{t} = \Big\{ \tilde{\mathcal{Y}}^{t+1}_{f,n} | \text{ action is not masked}, f, n \in \boldsymbol{\mathbbm{F}}_{s}, \boldsymbol{\mathbbm{N}} \Big\} = \\
 &\! \Big\{ \! \tilde{\mathcal{Y}}^{t+1}_{f,n} 
 \left|
    \genfrac{}{}{0pt}{0}  
    {\scriptscriptstyle \hspace{-1em} \qquad \sum_{\boldsymbol{\mathbbm{R}}} \widecheck{\mathcal{I}}_{r,f}  \leq \sum_{\boldsymbol{\mathbbm{N}}}{(\tilde{\mathcal{Y}}^{t+1}_{f,n} \cdot \widehat{\mathcal{C}}_n)} }
    {\scriptscriptstyle \sum_{\boldsymbol{\mathbbm{R}}}{(\tilde{\mathcal{X}}_{r,f}^{t+1} \cdot \tilde{\mathcal{Y}}^{t+1}_{f,n})} > 0 }  
    \right.
    \quad f, n  \in \boldsymbol{\mathbbm{F}}_{s}, \boldsymbol{\mathbbm{N}}
\Big\}
\end{aligned} 
\label{placement_action}
\end{equation}
\vspace{-3mm}
\normalsize

\vspace{-5pt}
\footnotesize
\begin{equation}
\begin{aligned} 
     {R}_\text{\tiny{PL}}^{t} \! &= \!\! \sum_{\boldsymbol{\mathbbm{R}}} ({\!\prod_{\boldsymbol{\mathbbm{F}}_{s_r}}\!\!\tilde{\mathcal{X}}_{r,f}^{t+1}}) -  \alpha ( \!\!\!\!\! \sum_{\boldsymbol{\mathbbm{F}}_{s}, \boldsymbol{\mathbbm{N}}, \boldsymbol{\mathbbm{T}}}
    \!\!\!\!\! \tilde{\mathcal{Y}}^{t+1}_{f,n} \overline{\mathcal{E}}_n + \!\!\!\!\!\!\!\!\!\!\!\! \sum_{\boldsymbol{\mathbbm{L}}^{t+1}, \boldsymbol{\mathbbm{P}}^{t+1}, \boldsymbol{\mathbbm{R}}, \Delta_r} \!\!\!\!\!\!\!\!\!\!\!\!\!\! \overline{\xi}_l \mathcal{J}^{t+1}_{p,l} \overrightarrow{\mathcal{R}}^{t+1}_{r,p} )
\end{aligned}
\label{placement_reward}
\end{equation}
\vspace{-3mm}
\normalsize

\myHighlight{
Table~{\ref{table:subproblems_mapping}} details the subproblems addressed in PERFECT, highlighting their physical aspects, corresponding decision layers, and underlying technologies. The modular decomposition enables hierarchical decision-making across multiple time scales, effectively enhancing adaptability, energy efficiency, and QoS assurance in aerial-terrestrial vehicular 6G networks.
}

\vspace{-5pt}
\section{Performance Evaluation}\label{sec:results}
This section evaluates our proposed method's efficiency. It begins with a convergence analysis, examining the impact of hyperparameters on the algorithm's performance. Subsequently, PERFECT is benchmarked against baseline methods using diverse metrics, demonstrating its superiority.

\vspace{-5pt}
\subsection{Simulation settings}

The simulations are conducted within a vehicular edge–cloud continuum, where UAVs initially fly at a fixed height and are randomly distributed across the network area. Key simulation parameters are summarized in Table~\ref{table:simulation_parameter}, where parameters follow a uniform distribution $\mathcal{U}$. \myHighlight{The parameters ensure that energy consumption is physically modeled following the aerodynamic formulation in, while the wireless channel reflects environment-dependent attenuation following. This configuration enables a fair and realistic comparison between PERFECT and baseline frameworks under varying conditions.}
To model UE mobility and reflect realistic user dynamics, we employ \myHighlight{Simulation of Urban Mobility (SUMO)} \cite{SUMO2018}, a \myHighlight{microscopic vehicular mobility simulator} designed for large-scale network environments. Specifically, we consider a grid environment with bidirectional roads, \myHighlight{while SUMO provides realistic trajectories across different urban zones of Oulu City in Finland, including both city-center and suburban areas, allowing us to reflect diverse densities and mobility patterns.} UEs follow the Manhattan mobility model, moving straight with a 50\% probability and turning left or right with a 25\% probability at intersections. \secH{The mobility model is inspired by the Manhattan-like urban cities and implemented with additional stochasticity in vehicle interactions and departure processes, providing diverse and generalized dynamics.} During the simulation, there are 50 UEs in the network, each capable of \myHighlight{generating various types of service requests with heterogeneous data rate requirements}, providing a practical context for assessing our proposed method.

\renewcommand{\arraystretch}{1.35}
\begin{table}[t!]
\caption{Simulation Parameters.}
\vspace{-10pt}
\label{table:simulation_parameter}
\begin{tabular}{C{0.83cm}C{4.2cm}C{2.6cm}}
Domain                      & Parameter           & Value \\ \hline
\hline
\multirow{6}{*}{Network}    & Node Processing Capacity ($\widehat{\mathcal{C}}_n$)             & $\sim \mathcal{U}(25, 70)$ Mbps  \\ \cline{2-3} 
                            & Node Energy Capability ($\overline{\mathcal{E}}_n$)              & $\sim \mathcal{U}(12, 36)$ Hz  \\ \cline{2-3}
                            & Link Bandwidth Capacity ($\widehat{\mathcal{L}}_l$)              & $\sim \mathcal{U}(10, 30)$ Mbps  \\ \cline{2-3}
                            & Link Latency ($\mathcal{D}^{t}_{r,l}$)                           & $\sim \mathcal{U}(4, 16)$  Mbps  \\ \cline{2-3}
                            & Link Energy Consumption ($\overline{\xi}_l$)             & $\sim \mathcal{U}(5, 8)$   Hz  \\ \cline{2-3}
                            & \myHighlight{UAV Weight, velocity ($W_n$, $V_w(t)$)} &\!\!\myHighlight{$ \mathcal{U}(4,6)$kg, \!$\mathcal{U}(8,12)$m/s}\\  \hline
\multirow{4}{*}{Area}       & Areas ($\boldsymbol{\mathbbm{A}}$) & 4*4 Grid                                     \\ \cline{2-3} 
                            & \myHighlight{Induced \!Power,\! Drag Coefficient ($I$,$\varsigma$)} & \myHighlight{0.08, 0.05} \\ \cline{2-3} 
                            & \myHighlight{Air Density ($\varphi$)} & \myHighlight{1.225 $kg.m^{-3}$} \\ \cline{2-3} 
                            & \myHighlight{Rotor Disk, Frontal Area ($\upsilon_r, \upsilon_f$)} & \myHighlight{0.6, 0.25 $m^2$} \\ \hline
\multirow{4}{*}{Service}    & Number of Composed Services ($\boldsymbol{\mathbbm{S}}$)         & 20                            \\ \cline{2-3} 
                            & Number of Atomic Functions ($\boldsymbol{\mathbbm{F}}_s$)        & 32                            \\ \cline{2-3}
                            & Service Duration ($\overrightarrow{\mathcal{T}_s}$)             & $\sim \mathcal{U}(3, 10)$ frames    \\ \cline{2-3} 
                            & Service Required Time Slots ($\widecheck{\mathpzc{T}}_r$)              & $\sim \mathcal{U}(3, 9)$  slots    \\ \hline 
                                     
\multirow{4}{*}{UE}         & Vehicular UEs ($\boldsymbol{\mathbbm{U}}$)                       & 50   \\ \cline{2-3} 
                            & Bandwidth Requirement ($\widecheck{\mathcal{L}}_{r}$)        & $\sim \mathcal{U}(2, 8)$ Mbps      \\ \cline{2-3} 
                            & Capacity Requirement ($\widecheck{\mathcal{I}}_{r,f}$)       & $\sim \mathcal{U}(8, 20)$ Mbps     \\ \cline{2-3} 
                            & Latency Requirement ($\widecheck{\mathcal{D}}_{r}$)                  & $\sim \mathcal{U}(50, 100)$ ms    \\ \hline

\multirow{8}{*}{Channel}    & Time slots per time frame ($\mathfrak{T}_t$)                    & 10                             \\ \cline{2-3}
                            & LoS Probability ($\theta_{a}^{\text{\tiny{LoS}}}$)                    & $\sim \mathcal{U}(0.2, 0.8)$   \\ \cline{2-3}
                            & \myHighlight{Rayleigh coefficient} (${{R}}^{\tau}_{c,a}$)         & $\sim \!\! \mathcal{R}(\sigma \!\! \sim \! \mathcal{U}(0.2, 0.8))$\\ \cline{2-3}
                            & Quality Threshold ($\widehat{\mathcal{Q}}$)                     & 1  \\ \cline{2-3}
                            & Channel Energy Consumption ($\overline{\mathcal{M}}_c$)          & $\sim \mathcal{U}(2, 8)$ Hz    \\ \cline{2-3}
                            & \myHighlight{Weather Attenuation ($\zeta(\Omega_{\tau})$)} & \myHighlight{$\{0,2.5,5.0\}$ dB} \\ \cline{2-3}
                            & \myHighlight{Shadowing Std. Deviation ($\eta_s(\Omega_{\tau})$)} & \myHighlight{$\{2.0,3.0,4.5\}$ dB} \\ \hline
\multirow{4}{*}{HDRL}       & Running episodes & 100,000 \\ \cline{2-3}
                            & Replay Memory ($\psi_{\text{\tiny{TP}}}, \psi_{\text{\tiny{PL}}}$)         & 2000, 1000                     \\ \cline{2-3}
                            & Discount Factor ($\Gamma$)                                       & 0.8                            \\ \cline{2-3}
                            & Scaling Factors ($\alpha$, $\chi$, $\kappa$)                     & 0.001, 0.5, 0.8                \\ \cline{2-3}
                            & \textit{EpsilonGreedy} process ($\epsilon$, $\epsilon'$, $\widetilde{\epsilon}$) & 1, 0.00005, 0.0001 \\ \hline
\multirow{4}{*}{TP}         & NN Layers (LSTM / Two CNN (kernel size, stride, and pooling size) / Three Fully Connected) & 128 units / (3, 2, 2) / (256, 128, 64 units) \\ \cline{2-3}
                            & Activation Function & Hyperbolic Tangent \\ \hline
\multirow{2}{*}{PL}         & NN Layers (Three Fully Connected) & (512, 256, 128 units) \\ \cline{2-3}
                            & Activation Function & Leaky ReLU \\ \hline
\end{tabular}
\begin{tablenotes}
\footnotesize
\item $\mathcal{U}$: Uniform distribution, $\mathcal{R}$: Rayleigh distribution, Mbps: Megabits per second, ms: millisecond, Hz: Hertz, ReLU: Rectiﬁed Linear Units
\end{tablenotes}
\vspace{-5pt}
\end{table}

\vspace{-7pt}
\subsection{Convergence Analysis}
\begin{figure}[t!]\centering
\vspace{-1pt}
\includegraphics[width=3.32in]{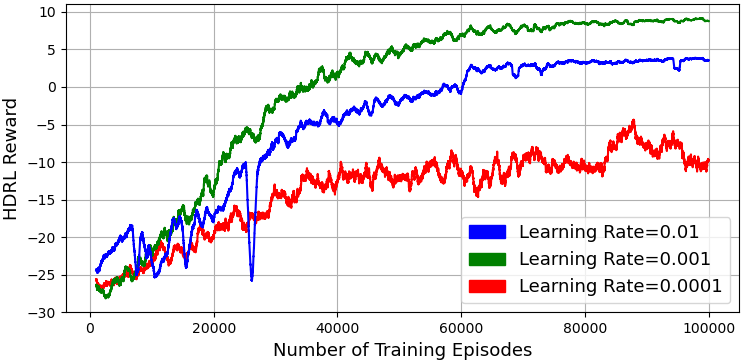}
\vspace{-6pt}
  \caption{Convergence performance of the PERFECT algorithm for different learning rates, highlighting its stability and reward optimization.}
    \vspace{-6pt}
    \label{figure:converge_lr}
\end{figure}

\begin{figure}[t!]\centering
\includegraphics[width=3.35in]{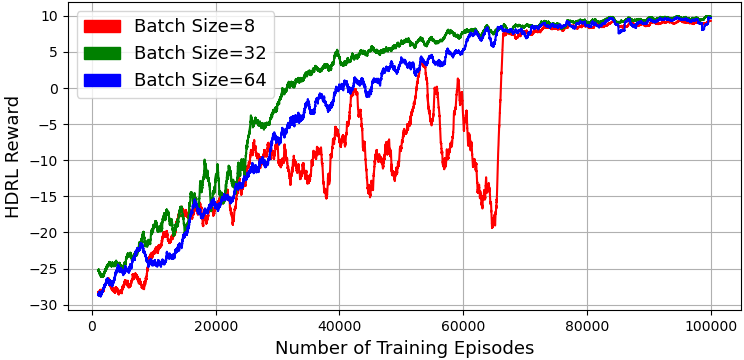}
\vspace{-6pt}
  \caption{Convergence performance of the PERFECT algorithm for different batch sizes, showing the trade-offs between stability and convergence speed.}
    \vspace{-6pt}
    \label{figure:converge_batch}
\end{figure}

We conduct experiments with varying hyperparameters to assess the proposed method's convergence behavior. The parameters are critical for training efficiency and stability, with the learning rate controlling the magnitude of NN weight updates and batch size defining sample numbers per training episode. Small learning rates result in prolonged training durations or even non-convergence because of ineffective loss minimization. Conversely, large rates lead to unstable training dynamics and prevent convergence, where rapid weight updates may overshoot optimal values or become trapped in local optima. Fig.~\ref{figure:converge_lr} demonstrates that a learning rate of 0.001 achieves optimal convergence, offering high rewards and stability across training episodes. Also, smaller batch sizes introduce variability and instability due to random sampling, while larger batch sizes increase overhead and slow convergence. As shown in Fig.~\ref{figure:converge_batch}, a batch size of 32 strikes the right balance, ensuring faster and smoother convergence, compared to the more variable results with a batch size of 8 and slower convergence with a batch size of 64. A noteworthy feature of our HDRL framework is that it considers two high- and low-level policies, leading to reward escalation once we start a high-level one (episode 25,000).

\vspace{-5pt}
\subsection{Comparison experiments}
We compare our proposed method against baseline approaches to demonstrate its effectiveness. The compared approaches include the optimal solution of ALLOCATE derived via CPLEX (with complete knowledge); a random selection strategy for UAV trajectory planning, channel selection, and service deployment; the Successive Convex Approximation-based (SCA) method \cite{gupta2023trajectory} that considers multi-UAV trajectory and resource planning; and the Hungarian and DDQN-based (HaDDQN) method \cite{qi2022energy}, which uses a DDQN approach for service placement and a Hungarian algorithm for RB allocation in co-channel settings. The SCA method is modified to include latency constraints instead of UAV recharging as considered in the original study. Likewise, the HaDDQN method is adapted to align with our framework by embracing the Hungarian algorithm for UAV trajectory decisions.

Three quantitative metrics critical to vehicular networks are considered to compare methods. First, the number of accepted requests (request coverage) indicates scalability in dynamic environments, essential for connectivity. Second, energy consumption quantifies the energy required for UAV movement, communication through channels, and resource utilization, addressing sustainability goals and operability extension. 
Third, E2E latency assesses the ability to ensure seamless user experiences, which is crucial for futuristic latency-sensitive applications like autonomous driving. These metrics collectively provide a comprehensive assessment of the method's effectiveness, highlighting the potential to balance service delivery, energy use, and E2E latency requirements.

To evaluate the different aspects of the problem and their implications for service orchestration, \myHighlight{we increase the number of requests, network nodes, and communication channels that are varied across simulation scenarios.} The first one assesses the response to varying UE and active requests to verify the ability to maintain high request acceptance under dynamic workload conditions. It simulates real-world request fluctuations, as seen in futuristic applications like intelligent transportation systems \cite{Oladimeji2023}, where vehicles generate variable real-time data during peak hours or emergencies. Second, varying network node and \myHighlight{areas through SUMO}\footnote{\myHighlight{The transition from sparse suburban to dense urban traffic is captured.}} evaluate the effect of network and \myHighlight{infrastructure} scale. The scenario reflects expected growth in 6G infrastructure size \cite{taleb20226g} to ensure effective performance across diverse environments, from small-scale edge-cloud systems to expansive, distributed networks. In futuristic networks, densely populated urban areas with high vehicular density and connectivity demand can strain channel availability \cite{ChannelConstraint}. Thus, we consider varying communication channel $\boldsymbol{\mathbbm{C}}$ as the third scenario that determines adaptability to constrained channel availability across diverse conditions.

\begin{figure*}[t!]\centering
\includegraphics[width=7.15in]{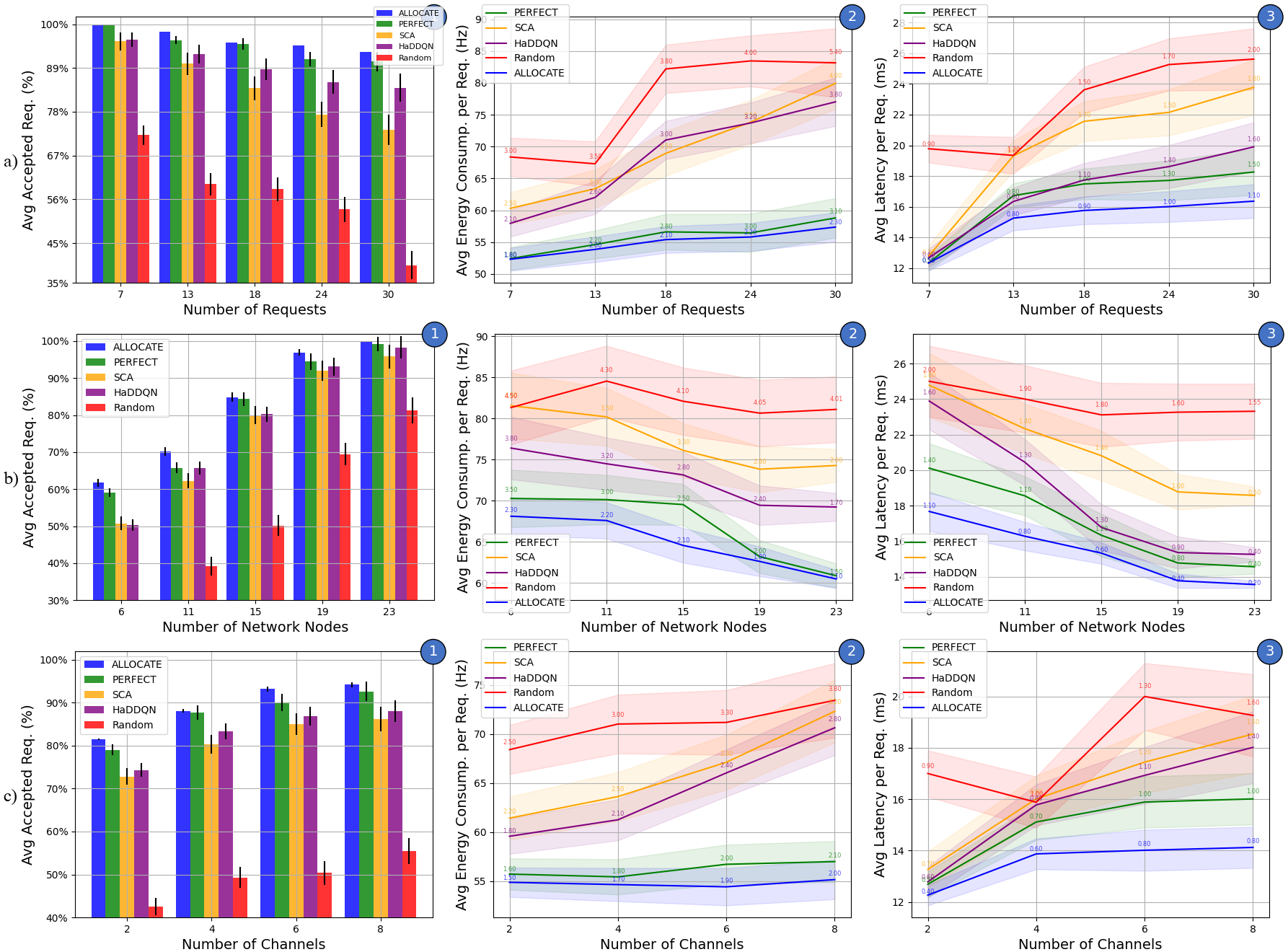}
\vspace{-12pt}
  \caption{(1) supported requests percentage, (2) energy consumption, and (3) E2E latency are compared between the ALLOCATE, PERFECT, SCA, HaDDQL, and random methods as (a) request set, (b) network size, and (c) channel size expand. \secH{The shaded regions indicate the standard deviation across multiple runs.}}
    \vspace{-5pt}
    \label{figure:output1}
\end{figure*}

Fig.~\ref{figure:output1} illustrates PERFECT and baseline methods comparisons in terms of accepted requests percentage (1) as well as energy consumption (2) and E2E latency (3) incurred by each request under increasing requests (a), nodes (b), and channels (c) scenarios. Notably, requests failing latency requirements are considered unacceptable, reflecting their inability to provide a satisfactory user experience. To ensure statistical robustness in a setting containing uniform distributions, we consider average values from multiple system runs, each using identical seed values for all methods. This approach ensures that each method operates under the same conditions in each iteration. In this regard, even in optimal solutions, energy consumption and E2E latency fluctuate due to dynamic factors like node/link capacities and evolving bandwidth requirements. Besides, the shaded regions around the trend lines represent standard deviations, indicating performance variability. \secH{The observed differences in shaded regions stem from the varying robustness of the methods to network dynamics, with the Random method showing the highest variability due to uninformed decisions, while PERFECT and ALLOCATE achieve more stable performance through adaptive allocation.}

\textbf{Scenario I:}
This scenario begins with increasing request numbers from 7 (minimal load) to 30 (extreme load) per time frame while keeping network size (10 nodes) and channel availability (10 channels) constant. \myHighlight{Fig.~{\ref{figure:output1}}(a) demonstrates the scalability of the proposed framework under different request numbers}. All methods exhibit a slight decline in accepted requests as the number of requests increases, as the number of network nodes remains fixed. Higher request volumes lead to increased E2E latencies due to diverse requests with varying requirements across networks with high-capacity nodes located far from PoAs. Also, to accommodate high demand, methods are compelled to deploy additional functions and utilize more links for transmitting requests and responses, thereby increasing energy consumption.

PERFECT maintains high request coverage and relatively stable energy consumption, outperforming HaDDQN, SCA, and random methods. Regarding E2E latency, both HaDDQN and PERFECT perform well, as they prioritize maintaining latency within acceptable requirements. The random method's poor request handling leads to service interruptions, latency infringement, and high energy consumption, rendering it unsuitable for real-world applications. SCA and HaDDQN's limited ability to predict requests contributes to their reduced performance, particularly in larger request volumes. Also, their TP methods overlook total energy minimization (discussed in Section~\ref{serviceOrchestrationPhase}), incurring higher energy consumption. SCA's lack of shared channel support further diminishes its acceptance range compared to HaDDQN, despite its comparable performance under lighter loads. HaDDQN, benefiting from its learning PL, achieves superior energy efficiency than SCA in function deployment, though it remains less efficient than PERFECT. PERFECT's predictive capabilities and Bayesian-based channel sharing, combined with efficient TP and PL modules, ensure E2E latency compliance. It strategically deploys functions farther from UEs to reduce energy consumption, resulting in slightly higher, but within latency tolerances in some cases. The gap between ALLOCATE and PERFECT is minimal and stems from occasional prediction errors that necessitate deploying slightly more functions. However, these increases are negligible in practical scenarios and underscore PERFECT's efficient prediction algorithm.
\myHighlight{Such consistency demonstrates the predictive module's adaptability to unexpected mobility or traffic patterns, as it dynamically updates its knowledge base to reflect new behavior, preventing service degradation and sustaining stable orchestration efficiency.}

\textbf{Scenario II:}
The second scenario scales the network infrastructure from 6 to 23, with fixed requests (15) and channels (10). As the network size increases, the number of areas also expands (from 4×4 to 6×6). \myHighlight{Fig.~{\ref{figure:output1}}(b) shows the scalability and flexibility of the proposed framework in varying environments and under different infrastructure.} Initially, all methods show lower acceptance rates due to an imbalance between requests and limited UAVs. As network density increases, acceptance rates improve with additional network resources. Energy consumption starts high due to reliance on high energy-intensive edge nodes but gradually decreases as UAVs are deployed, requiring fewer links for service delivery. Similarly, E2E latency declines as UAVs are positioned closer to UEs, reducing transmission latencies.

The proposed method exhibits strong scalability and efficient energy consumption while performing practically similarly to HaDDQN regarding latency. The random approach's energy-intensive node selection leads to high energy consumption and prolonged latency despite minor improvement due to increased nodes. As seen in the first scenario, SCA and HaDDQN struggle with request prediction, which affects their performance in smaller networks. However, as the network size increases, the acceptance of both methods increases since there are sufficient nodes in each area. SCA struggles with high latency in limited networks due to distant function deployments, occasionally violating latency requirements. As the network size increases, SCA's latency improves, reflecting its ability to better utilize the expanded infrastructure. Energy consumption in PERFECT is notably efficient, with lower UAV movement and deployment energy compared to SCA and HaDDQN, owing to its HDRL algorithm that optimally selects proper nodes for service coverage and delivery. The HDRL algorithm enhances this efficiency through high-level policies that become increasingly impactful as the network size grows. While HaDDQN and PERFECT exhibit similar energy usage in terms of deployment, PERFECT's predictive capabilities offer minor improvements. As UAV numbers grow, PERFECT further reduces energy consumption by accurately predicting user behavior and optimizing unnecessary UAV movements.
\myHighlight{The information gathering DRL agents continuously capture mobility-induced variations, enabling PERFECT to anticipate real-time mobility and adjust TP decisions.}
This optimization also lowers E2E latency, contributing to the method's overall effectiveness in handling scalability. ALLOCATE performs comparably, especially in the presence of sufficient resources in the network, with minor variations due to slight prediction errors during information gathering.

\textbf{Scenario III:}
In this scenario, the \myHighlight{number of channels} is increased from 2 to 8 while keeping the number of requests (10) and network nodes (10) fixed. Channel availability influences system performance, with noticeable improvements observed when the number of channels exceeds two. This enhancement is attributed to the requirement of three or more time slots for most requests. With more channels, accepted requests increase across all methods due to additional available RBs for transmission. However, this growth raises energy consumption because of the unique energy demands of each channel and the complexity of managing shared channels. Similarly, the rise in accepted requests slightly increases E2E latency, reflecting the higher system load.

In this scenario, PERFECT achieves higher request acceptance rates and superior energy efficiency than SCA and HaDDQN due to its Bayesian algorithm for channel quality prediction that optimizes channel allocation. HaDDQN lacks channel prediction, leading to inefficient utilization and high energy consumption. Furthermore, the absence of support for shared channels led to an increase in channel energy consumption and E2E latency for SCA. The difference in accepted requests between ALLOCATE and PERFECT stems from channel quality prediction errors and the proposed greedy channel selection. E2E latency remains stable for PERFECT across channel configurations, whereas other methods, especially SCA and random, experience rising latency with increasing channel availability, highlighting their limitations in managing higher channel counts.


\begin{figure*}[t!]\centering
\vspace{-10pt}
\includegraphics[width=7.1in]{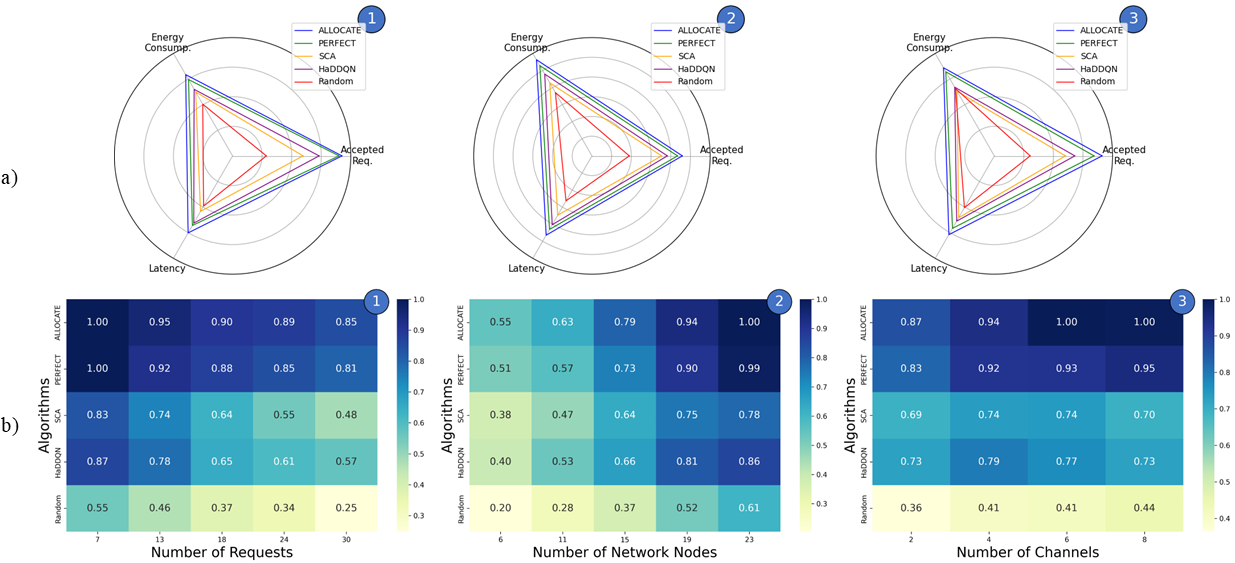}
\vspace{-12pt}
  \caption{Performance metrics comparison of the baseline methods as (1) request set, (2) network size, and (3) channel size expand. \myHighlight{In (a), radar plots present the normalized trade-offs among accepted requests, latency, and energy consumption, demonstrating how each algorithm behaves under varying scenarios. In (b), the heat maps depict the ratio of accepted requests to energy consumption (as objectives) under the same conditions}.}
    \label{figure:output2}
    \vspace{-6pt}
\end{figure*}

\vspace{-5pt}
\subsection{Discussion}

A comparison of PERFECT with alternative approaches highlights its superior timely response. On average, PERFECT produces results 88\% faster than the optimal approach. This speed advantage is supported by its complexity analysis: $\boldsymbol{\mathbbm{T}}\big( (\boldsymbol{\mathbbm{U}} \!+\! \boldsymbol{\mathbbm{R}})\boldsymbol{\mathbbm{A}}) \!+\! (\boldsymbol{\mathbbm{N}}_{\text{\tiny{UAV}}} \boldsymbol{\mathbbm{A}}) \!+\! (\boldsymbol{\mathbbm{N}} \boldsymbol{\mathbbm{U}} \boldsymbol{\mathbbm{C}} \mathfrak{T}_t) \!+\! (\boldsymbol{\mathbbm{N}} \boldsymbol{\mathbbm{F}}_s \!+\! \boldsymbol{\mathbbm{P}} \boldsymbol{\mathbbm{U}})\big)$. This complexity arises from \myHighlight{(i) evaluating $\boldsymbol{\mathbbm{U}}$ user trajectories and $\boldsymbol{\mathbbm{R}}$ requests over areas (Prediction); (ii)} evaluating each UAV in each area (TP)\myHighlight{; (iii)} assessing nodes, UEs, channels, and RBs (MAC)\myHighlight{; and (iv)} analyzing nodes and functions for function deployment, and determining optimal paths for requests (PL). 
\myHighlight{The increased number of nodes and requests affects the runtime slightly, but this marginal cost is offset by performance gains in acceptance, latency, and energy, as demonstrated in Fig.~{\ref{figure:output1}}.
Notably, PERFECT's complexity is dominated by lightweight online inference, while expensive offline training is amortized over long-term system usage. The HDRL training stage is computationally demanding due to iterative exploration, reward evaluation, and parameter updates within multiple simulated episodes; however, this cost is incurred once before deployment. In contrast, the online inference phase requires sub-second execution times on standard edge hardware.
Hence, PERFECT achieves an effective balance between computational overhead and performance metrics, confirming scalability and HDRL design efficiency in next-generation vehicular networks.}

\myHighlight{
To benchmark complexity, PERFECT was compared with SCA and HaDDQN. In SCA, the non-convex problem is decomposed into sequential convex subproblems, each solved via an interior-point method over all nodes, areas, users, and channels, yielding {$\boldsymbol{\mathbbm{T}} (\boldsymbol{\mathbbm{N}} \boldsymbol{\mathbbm{A}} \boldsymbol{\mathbbm{U}} \boldsymbol{\mathbbm{P}} \boldsymbol{\mathbbm{C}} \mathfrak{T}_t)^{3.5}$}, complexity. While SCA converges to a stationary point, its runtime scales poorly with network size and channel numbers. HaDDQN incurs a per-frame cost of \mbox{$\boldsymbol{\mathbbm{T}} \big(( (\boldsymbol{\mathbbm{U}} \!+\! \boldsymbol{\mathbbm{R}}) \boldsymbol{\mathbbm{P}} \boldsymbol{\mathbbm{C}} \mathfrak{T}_t)^3 \!+\! (\boldsymbol{\mathbbm{N}}_{\text{\tiny{UAV}}} \boldsymbol{\mathbbm{A}})\big)$}, dominated by the cubic complexity of the assignment operations, which relies on the Hungarian algorithm for user-resource and UAV-area association, followed by a DDQN-based decision. Thus, PERFECT maintains lower practical complexity and faster execution. As shown in Fig.~{\ref{figure:output2}}.a, PERFECT consistently surpasses SCA and HaDDQN in the oracle's objective, owing to its prediction-aware decisions, Bayesian channel adaptation, and hierarchical coordination. Each radar in this figure is generated by first normalizing (min-max) the averaged values of accepted requests, energy consumption, and E2E latency collected across multiple scenarios at different densities, and then plotting the aggregated triplet for each algorithm to visualize its overall performance balance in each scenario.
}

\myHighlight{Fig.~{\ref{figure:output2}}.a) presents heat maps} of accepted requests and energy consumption, highlighting balanced behavior across different metrics. PERFECT achieves over 92\% of ALLOCATE's optimal rates for varying requests, significantly outperforming the random approach--which drops to 24\% at higher request levels. In the network scaling scenario, PERFECT's performance rates rise from 51\% to 99\%, achieving 93\% of ALLOCATE's performance. Similarly, for channel variations, PERFECT excels with a 92\% of ALLOCATE, outperforming SCA and HaDDQN, which plateau at 71-74\%. 
\myHighlight{Besides, PERFECT's performance metrics remain stable when the number of users, networks, and channels increases, as depicted in Fig.~{\ref{figure:output2}}.b). 
Each heatmap is constructed by computing the ratio of total accepted requests to total energy consumption based on averaged simulation statistics for every method that provides a compact representation of the objective trade-off. The illustrations of the PERFECT evaluations underscore the proposed method's scalability and ability to (i) balance competing objectives effectively, (ii) sustain efficient operation, and (iii) effectively adapt to varying network conditions.
}

\section{Conclusion}\label{sec:conclusion}
In this study, we proposed a comprehensive framework for composed service orchestration in 6G aerial-terrestrial networks, addressing the intertwined challenges of mobility, resource planning, and service coverage. Our orchestration approach ensures that we are able to meet the diverse requirements of modern vehicular applications, resulting in efficient resource allocation and high-quality service provisioning. An MINLP problem of service orchestration in an integrated aerial-terrestrial network, while accounting for capacity constraints, changing user behavior, and E2E latencies, was first formulated to maximize service coverage while optimizing energy consumption. To solve the NP-hard problem, the integration of HDRL with predictive modeling enabled efficient UAV trajectory planning and resource-efficient service placement, ensuring QoS compliance and enhanced system performance. Our simulation results highlighted significant improvements in request acceptance, energy efficiency, and latency minimization, outperforming traditional and state-of-the-art methods. This framework underscores the transformative potential of HDRL-driven solutions for managing the complexity and scalability of next-generation vehicular networks. 

Future works focus on further enhancing the scalability of the proposed method. Promising directions include integrating federated learning to enable decentralized training and decision-making, and exploring multi-agent RL to improve coordination among UAVs and network nodes \myHighlight{in heterogeneous, multi-tiered environments involving satellite, optical wireless, and radio frequency communication nodes.} Also, we plan to explore opportunities for using Large Language Models \cite{LLMRA} for high-level decisions to assist in \myHighlight{reasoning about resource allocation challenges in emerging quantum internet settings, such as qubit routing and link-level scheduling} \cite{quantumInternet}.

\section*{Acknowledgment}
The work in this paper was supported in part by the Federal Ministry of Research, Technology, and Space (BMFTR), Germany, through the Project 6GEM+ under Grant 16KIS2411; and in part by the 6G-Path project (Grant No. 101139172).

\bibliographystyle{IEEEtran}
\bibliography{IEEEabrv, conf_short, main}

\end{document}